\documentclass[draftcls,12pt,onecolumn]{IEEEtran}
\usepackage{cite}
\usepackage[pdftex]{graphicx}
\graphicspath{{figure/}}
\usepackage{array}
\usepackage{subfigure}
\usepackage{stfloats}
\usepackage{url}
\usepackage{nicefrac}
\usepackage{amssymb}
\usepackage{amsmath}
\usepackage{amsfonts}
\usepackage{amssymb}
\usepackage{float}
\usepackage{color}
\usepackage{kbordermatrix}

\floatstyle{ruled}
\newfloat{algorithm}{tbp}{loa}
\floatname{algorithm}{Algorithm}

\newtheorem{proposition}{Proposition}
\newtheorem{definition}{Definition}

\newcommand{\abs}[1]{\left\lvert#1\right\rvert}

\def\b#1{\mathbf{#1}}
\def\t#1{\text{#1}}

\begin{document}

\title{On Secrecy Rate Analysis of MIMO Wiretap Channels Driven by Finite-Alphabet Input}
\author{Shafi Bashar, {\em Student Member, IEEE}, Zhi Ding, {\em Fellow, IEEE}, and Chengshan Xiao {\em Fellow, IEEE}
\thanks{S. Bashar and Z. Ding are with the Department of Electrical and Computer Engineering, Univ. of California, Davis, CA, 95616, USA. e-mail: \{shafiab, ding\}@ece.ucdavis.edu.}
\thanks{C. Xiao
is with the Department of Electrical and Computer Engineering, Missouri University of Science and Technology,
Rolla, MO, 65409, USA. e-mail: xiaoc@mst.edu.
}\thanks{This material is based upon work supported by the National Science Foundation under Grants No. 0520126.
}}
\markboth{Submitted to IEEE Transactions on Communications, April 4, 2011}{}

\maketitle

%%%%%%%%%%%%%%%%%%%%%%%%%%%%%%%%%%%%%%%%%%%%%%%%%%%%%%%%%
\begin{abstract}
This work investigates the effect of finite-alphabet input constraint on the secrecy rate of
a multi-antenna wiretap channel.
Existing works have characterized maximum achievable secrecy rate or
secrecy capacity for single and multiple antenna systems based on
Gaussian source signals and secrecy code.
Despite the impracticality of Gaussian sources,  the compact closed-form expression of
mutual information between linear channel Gaussian input and corresponding output
has led to broad use of the Gaussian input assumption in physical secrecy analysis.
For practical considerations, we study the effect of finite discrete-constellation on the achievable secrecy 
rate  of multiple-antenna wire-tap channels.  Our proposed precoding scheme converts the underlying multi-antenna 
system into a bank of parallel channels. Based on this precoding strategy, we develop a decentralized power allocation 
algorithm based on dual decomposition to maximize the achievable secrecy rate. In addition, 
we analyze the achievable secrecy rate for finite-alphabet inputs in low and high SNR regions.
Our results demonstrate substantial difference in secrecy rate between systems
given finite-alphabet inputs and systems with Gaussian inputs.
\end{abstract}
%%%%%%%%%%%%%%%%%%%%%%%%%%%%%%%%%%%%%%%%%%%%%%%%%%%%%%%%%


%%%%%%%%%%%%%%%%%%%%%%%%%%%%%%%%%%%%%%%%%%%%%%%%%%%%%%%%%%
\section{Introduction}
\label{sec:introduction}
\IEEEPARstart{W}{ireless} communications, with increasing coverage
and applications, are vulnerable to potential
security compromises such as passive eavesdropping and active jamming.
Traditionally, network planners have relegated system security considerations to higher network layers of the OSI protocol stack 
through authentication and cryptography.
However, in recent years, there have been growing research interests in the
security analysis of wireless systems
from a physical layer and information theoretic perspective.
In a wiretap channel environment originally introduced by Wyner \cite{wyner1975wire}, a
sender ``Alice'' wishes to transmit a secret message to the intended receiver
``Bob'' in the presence of a passive eavesdropper ``Eve''. Wyner \cite{wyner1975wire} showed that when the Alice-to-Eve channel is degraded from the Alice-to-Bob channel,
Alice can encode and send secure messages to the destination at a non-zero secrecy rate.
In \cite{csiszar1978broadcast}, a generalization for the non-degraded
broadcast channel is proposed, and in \cite{hellman1978gaussian}, secrecy capacity of a
Gaussian wiretap channel is shown to be achievable by adopting a random Gaussian codebook.
In \cite{khisti_MIMOME}, the secrecy capacity of a multi-antenna Gaussian wiretap
channel is shown to be achievable using a suitable input covariance matrix and by
encoding the message using a Gaussian random codebook.

For both single- and multi-antenna Gaussian wiretap channels, the codebook that achieves secrecy capacity turns out to be Gaussian.
However, such codebooks are not implementable in practice. In real world systems,  input codebook consists of finite set of equi-probable 
constellation points (e.g. $M$-QAM, $M$-PAM etc.). Therefore, in contrast to the Gaussian codebook, 
practical wiretap codes must consist of finite-alphabet symbols. Because of this constraint, 
the achievable secrecy rate for a finite-alphabet input scenario would differ from the secrecy rate achieved by a Gaussian codebook.

A recent work \cite{rodrigues2010gaussian} considered the effect of $M$-PAM input on the secrecy rate of a Gaussian 
wiretap channel and provided the necessary condition for power allocation to maximize the achievable secrecy rate. 
In \cite{rodrigues2010gaussian}, results were also extended to the case of parallel Gaussian wiretap channels. 
In \cite {basharsecrecy}, we have investigated the effect of finite-alphabet input on the ergodic secrecy rate of a multiple-input single-output and single-eavesdropper (MISOSE) 
system. To continue our progresses in this work, we investigate the effect of finite-alphabet input in a
more general setting of a multiple-input  multiple-output and multiple-eavesdropper (MIMOME) system. 
The specific contributions of the work are summarized below :

\begin{itemize}
\item In order to quantify the effect of finite-alphabet input on MIMOME systems, we propose
the application of precoding matrix to transform the MIMOME channel into a bank of parallel channels. 

\item We propose a power allocation optimization framework based on decentralized dual decomposition 
technique to maximize the achievable secrecy rate of MIMOME systems with an arbitrary but known input distribution. 

\item We provide secrecy rate analysis of MIMOME systems with finite-alphabet inputs at
low and high SNR regions. Our findings suggest that similar to the Gaussian wiretap channel, proper transmission power 
should be diverted at high signal-to-noise ratio (SNR) in case of finite-alphabet input albeit with different effect. 
\end{itemize}

We organize the rest of the paper as follows. We begin with the system model in section \ref{sec:system_model}. 
In section \ref{sec:secrecy_rate_problem} we propose a  linear precoding scheme that transforms the MIMOME wiretap channel into a 
set of parallel channels. Based on this precoding scheme, we reformulate the secrecy rate problem for an arbitrary input distribution. 
In section \ref{sec:power_allocation}, we develop a decentralized power allocation algorithm based
on dual decomposition that maximizes the achievable secrecy rate for an arbitrary distribution. In section \ref{sec:analysis}, we 
further consider the special case of Gaussian input and present a modified water-filling power allocation strategy 
by considering the secrecy constraint. We then extend the modified water-filling power allocation scheme to analyze 
the secrecy rate for an arbitrary input distribution in both low and high SNR regions. 
In section \ref{sec:numerical_result} we present numerical results before concluding with section \ref{sec:conclusion}.

%%%%%%%%%%%%%%%%%%%%%%%%%%%%%%%%%%%%%%%%%%%%%%%%%%%%%%%%%%

%%%%%%%%%%%%%%%%%%%%%%%%%%%%%%%%%%%%%%%%%%%%%%%%%%%%%%%%%%
\section{Preliminaries and System Description}
\label{sec:system_model}

Throughout this work, we use notations $\t{tr}(.)$, $\t{det}(.)$ 
and superscript $\{\cdot\}^H$,  respectively,
to denote the trace, the determinant, and the conjugate transpose of a matrix.

\subsection{System Model}

We consider a MIMO (multiple-input multiple-output) wiretap system model in which the transmitter
(Alice), the intended receiver (Bob), and the passive eavesdropper (Eve),
respectively, have $m_a$ (transmit), $m_b$ (receive), and $m_e$ (eavesdrop) antennas.
Denote the received signals at Bob and Eve as $\b y_b$ and $\b y_e$, respectively.
Their received signals are written as
\begin{eqnarray}
\b y_{b} & = & \b H_{b}\b x+\b n_{b} \label{eq:bob}\\
\b y_{e} & = & \b H_{e}\b x+\b n_{e} \label{eq:eve}
\end{eqnarray}
where $\b H_{b}\in\mathbb{C}^{m_b\times m_a}$ and $\b H_{e}\in\mathbb{C}^{m_e \times m_a}$ denote, respectively, the flat-fading
MIMO channels, from
Alice-to-Bob and from Alice-to-Eve. The noise $\b n_b \in \mathbb{C}^{m_b}$ and $\b n_e\in\mathbb{C}^{m_e}$
are zero-mean identity matrix variance complex Gaussian random vectors independent of each other.
The data signal is $\b x \in \mathbb{C}^{m_a}$ transmitted by Alice in the form of $\b x = \b W \b s$,
in which $\b W$ is a linear precoding matrix.  We denote $\b s$ as a random vector with zero mean entries and identity correlation matrix. We constrain the total transmission power by a peak level $P_T$, i.e., $\t{tr}\{\b K_{\b x} \}\leq P_T$,
where $\b K_{\b x} = \mathbb{E} \{\b x \b x^H \}$ is the covariance matrix of the transmitted signal vector.

The secrecy capacity of the above system is achievable by using a Gaussian random
codebook \cite{khisti_MIMOME}.  First, let
$A\succeq B$ denote that $A-B$ is non-negative definite.
The secrecy capacity is the solution
 of the following optimization problem
\begin{equation}
\begin{split}
&\underset{\b K_{\b x}}{\t{maximize}}~\log\t{det}\left(\b I+\b H_{b}\b K_{\b x}\b H_{b}^{H}\right)-\log\t{det}\left(\b I+\b H_{e}\b K_{\b x}\b H_{e}^{H}\right)\\
&\t{subject\, to\,:}~~~~\b K_{\b x}\succeq0,\,\,{\b{K}_{\b x}=\b {K}_{\b x}^{H}}\\
&\phantom{\t{subject\, to\,:}~~~~}\t{tr}\left(\b K_{\b x}\right)\leq P_{T}.
\end{split}
\label{eq:main_problem}
\end{equation}

In order to realize an achievable secrecy rate for an arbitrary input signaling, we will generalize
the objective function. Instead of the optimal Gaussian signaling as used in the above optimization
problem, we replace the objective function with the more general form of $I\left(\b x;\b y_{b}\right)-I\left(\b x;\b y_{e}\right)$,
where $I\left(\b x;\b y\right)$ represents the mutual information between the input and output vectors $\b x$ and $\b y$.

The optimization problem described in \eqref{eq:main_problem} is a non-convex optimization problem (except for
the special case of $m_b = m_e = 1$ \cite{li2007secret,shafiee2007achievable}). Even for simple cases,
the objective function possesses a number of local maxima. Therefore, the optimum value of $\b K_x$ is in
general not known. In particular, for an arbitrary input signaling without any closed-form mutual information,
solving the above optimization problem can be a difficult task. Thus, instead of solving the above optimization problem,
we consider a particular linear precoding scheme similar to the one
provided in \cite{khisti_MIMOME}. The proposed linear precoding scheme first
transforms the MIMOME channel into a bank of parallel channels. 
This step allows us to gain a better understanding of the effect of finite-alphabet input on the secrecy of the system.
It enables us to gain better insights into future system implementation. Such linear precoding scheme is 
justifiable from practical system implementation perspective. In addition, 
at high SNR, such precoding scheme is known to achieve the capacity of a MIMOME system \cite{khisti_MIMOME}.

\subsection{Preliminaries}

\begin{definition}
Similar to \cite{khisti_MIMOME}, we define the following subspaces
\begin{eqnarray*}
\mathcal{S}_{b}    &=~\t{null}\left(\b H_{b}\right)^{\bot}&\cap~\,\,\t{null}\left(\b H_{e}\right)\\
\mathcal{S}_{b,e}  &=~\t{null}\left(\b H_{b}\right)^{\bot}&\cap~\,\,\t{null}\left(\b H_{e}\right)^{\bot}\\
\mathcal{S}_{e}   &=~\t{null}\left(\b H_{b}\right)&\cap~\,\,\t{null}\left(\b H_{e}\right)^{\bot}\\
\mathcal{S}_{n}   &=~\t{null}\left(\b H_{b}\right)&\cap~\,\,\t{null}\left(\b H_{e}\right).
\end{eqnarray*}
\hfill $\Box$\\
In fact,  subspace $\mathcal{S}_{b}$ corresponds to the class of input with non-zero
gain towards the direction of Bob only.
Subspace $\mathcal{S}_{b,e}$  corresponds to the class of input with non-zero
gain in the direction of both Bob and Eve.
$\mathcal{S}_{e}$ corresponds to the class of input with non-zero
gain in the direction of Eve only. Finally, $\mathcal{S}_{n}$ is
the subspace with non-zero gain in the direction occupied by
neither Bob nor Eve.
%
%and  are the
%subspaces corresponding to the classes of inputs with non-zero gain towards the direction
%of Bob only, both Bob and Eve, Eve only, neither Bob nor Eve respectively.
%
Define $k=\t{rank}\left(\left[\begin{array}{cc}
\b H_{b}^{H} & \b H_{e}^{H}\end{array}\right]^{H}\right)$ and hence $\t{dim}\left(\mathcal{S}_{n}\right)=m_{a}-k$.
In addition, we define, $r=\t{dim}\left(\mathcal{S}_{b}\right)$ and $s=\t{dim}\left(\mathcal{S}_{r,e}\right)$.
Therefore, $\t{dim}\left(\mathcal{S}_{e}\right)=k-r-s$.
\end{definition}

\begin{definition}
We recall the following definition of generalized singular value decomposition (GSVD) \cite{paige1981towards,khisti_MIMOME} that we will use for our analysis. The GSVD of the pair $\left(\b H_b, \b H_e\right)$ takes the following form

\[
\b H_{b}=\boldsymbol{\Psi}_{b}~\boldsymbol{\Sigma}_{b}
\kbordermatrix {~ & k & m_a-k \cr
		               ~ &\boldsymbol{\Omega}^{-1}  &  \b 0 \cr}
~\boldsymbol{\Psi}_{a}^{H}
\]
\[
\b H_{e}=\boldsymbol{\Psi}_{e}~\boldsymbol{\Sigma}_{e}
\kbordermatrix {~ & k & m_a-k \cr
		               ~ &\boldsymbol{\Omega}^{-1}  &  \b 0 \cr}
~\boldsymbol{\Psi}_{a}^{H}
\]
where $\boldsymbol{\Psi}_{a} \in \mathbb{C}^{m_a\times m_a}$, $\boldsymbol{\Psi}_{b} \in \mathbb{C}^{m_b\times m_b}$, and $\boldsymbol{\Psi}_{e} \in \mathbb{C}^{m_e\times m_e}$ are unitary matrices. $\boldsymbol{\Omega} \in \mathbb{C}^{k\times k}$ is a non-singular matrix. $\boldsymbol{\Sigma}_{b}\in\mathbb{C}^{m_b\times k}$ and $\boldsymbol{\Sigma}_{e}\in \mathbb{C}^{m_e\times k}$ have the following form

\[
\boldsymbol{\Sigma}_b = \kbordermatrix{~		   & k-r-s	 & s 		& r 		\cr
										m_b - r - s & \b 0 	& \b 0 		& \b 0 	\cr
										s                 & \b 0 	& \b D_b 	& \b 0 	\cr
										r                 & \b 0 	& \b 0     	& \b I 	\cr}
\]

\[
\boldsymbol{\Sigma}_e = \kbordermatrix{~		   & k-r-s	 & s 		& r 		\cr
										k - r - s      & \b I	 	& \b 0 		& \b 0 	\cr
										s                 & \b 0 	& \b D_e 	& \b 0 	\cr
										m_e-k+r    & \b 0 	& \b 0     	& \b 0	\cr}
\]

\noindent $\b D_b=\t{diag}\left(\left\{b_1,\ldots,b_s\right\}\right)$ and $\b D_e=\t{diag}\left(\left\{e_1,\ldots,e_s\right\}\right)$ are diagonal real matrices with dimension $s\times s$. The diagonal entries of $\b D_b$ and $\b D_e$ are arranged in the following orders:
\[0<b_1\leq\ldots b_s < 1\] \[1>e_1\geq \ldots \geq e_s > 0\] and \[b_i^2 + e_i^2 =1,~\mbox{for}~i=1,\ldots,s .\]
\hfill $\Box$
\end{definition}
%%%%%%%%%%%%%%%%%%%%%%%%%%%%%%%%%%%%%%%%%%%%%%%%%%%%%%%%%%

%%%%%%%%%%%%%%%%%%%%%%%%%%%%%%%%%%%%%%%%%%%%%%%%%%%%%%%%%%
\section{Secrecy Rate Problem Formulation}
\label{sec:secrecy_rate_problem}

\subsection{Proposed Precoding Strategy}
\label{sec:precoding_strategy}

Given the above two definitions in hand, we propose the following precoding matrix

\begin{equation}
\b W = \boldsymbol{\Psi}_{a}\b B\b P^{\nicefrac{1}{2}}
\label{eq:precoding_matrix}
\end{equation}
where, $\b B$ is defined as follows
\begin{equation}
\b B = \kbordermatrix{
~ 		& k 							& m_a - k 	\cr
k		& \boldsymbol{\Omega}		& \b 0		\cr
m_a-k	& \b 0						& \b 0 		\cr
}
\end{equation}
\noindent $P = \t{diag}\left(\left\{p_1,\ldots,p_{m_a}\right\}\right)$ is a diagonal power allocation matrix. The proposed precoding matrix in Eq. \eqref{eq:precoding_matrix} is similar to the precoding strategy defined in \cite{khisti_MIMOME}. However, unlike the precoding scheme in \cite{khisti_MIMOME}, the diagonal elemens $\left\{p_i\right\}$ in the power allocaion matrix can have different values. As will be evident in the subsequent sections, such differentiation is extremely important due to the finite nautre of the input constellations considered in this work.

By using the above precoding strategy, the system equations of \eqref{eq:bob}, \eqref{eq:eve} become
\begin{eqnarray}
\b y_{b}&=&\boldsymbol{\Psi}_{b}~\boldsymbol{\Sigma}_{b}
\kbordermatrix {~ & k     	&   m_a-k 	\cr
		                ~ & \b I 	&   \b 0 		\cr}
~\b P^{\nicefrac{1}{2}}~\b s+\b n_{b} \label{eq:bob_precoded}\\
\b y_{e}&=&\boldsymbol{\Psi}_{e}~\boldsymbol{\Sigma}_{e}
\kbordermatrix {~ & k     	&   m_a-k 	\cr
		                ~ & \b I 	&   \b 0 		\cr}
~\b P^{\nicefrac{1}{2}}~\b s+\b n_{e} . \label{eq:eve_precoded}
\end{eqnarray}

\noindent Pre-multiplying \eqref{eq:bob_precoded} and \eqref{eq:eve_precoded} with $\boldsymbol{\Psi}_{b}^H$ and $\boldsymbol{\Psi}_{e}^H$,
respectively, we have the following equivalent equations
\begin{eqnarray}
\tilde{\b y}_{b}&=&\tilde{\b H}_{b}\,\b s  +\tilde{\b n}_{b} \label{eq:bob_parallel}\\
\tilde{\b y}_{e}&=&\tilde{\b H}_{e}\,\b s  +\tilde{\b n}_{e} \label{eq:eve_parallel},
\end{eqnarray}
where we use the notations $\tilde{\b y}_{b} = \boldsymbol{\Psi}_{b}^H \b y_b$, $\tilde{\b y}_{e} = \boldsymbol{\Psi}_{e}^H \b y_e$, $\tilde{\b n}_{b} = \boldsymbol{\Psi}_{b}^H \b n_b$ and $\tilde{\b n}_{e} = \boldsymbol{\Psi}_{e}^H \b n_e$. The new equivalent
channel matrices $\tilde{\b H}_b = \boldsymbol{\Sigma}_{b}\left[\begin{array}{cc}
\b I & \b 0\end{array}\right]\b P^{\nicefrac{1}{2}}$ and  $\tilde{\b H}_e = \boldsymbol{\Sigma}_{e}\left[\begin{array}{cc}
\b I & \b 0\end{array}\right]\b P^{\nicefrac{1}{2}}$ are specified 
in Eqs. \eqref{eq:Hb_tilde} and \eqref{eq:He_tilde}.

\begin{figure*}[!t]
\begin{equation}
\tilde{\b H}_b  = \kbordermatrix{
~ 			& k-r-s						& s 								& r 			& m_a-k	\cr
m_b-r-s		& \b 0						& \b 0							& \b 0		& \b 0	\cr	
s			& \b 0						& \t{diag}\left(\left\{b_1\,\sqrt{p_{k-r-s+1}},\ldots,b_s\, \sqrt{p_{k-r}}\right\}\right)
																		& \b 0		& \b 0	\cr
k			& \b 0						& \b 0 							& \t{diag}\left(\left\{ \sqrt{p_{k-r+1}},\ldots,\sqrt{p_k} \right\}\right)		& \b 0	\cr
}
\label{eq:Hb_tilde}
\end{equation}
\begin{equation}
\tilde{\b H}_e = \kbordermatrix{
~			&k-r-s											&s		&m_a-k+r	\cr
k-r-s		&\t{diag}\left(\left\{\sqrt{p_{1}},\ldots, \sqrt{p_{k-r-s}}\right\}\right)	
															&\b 0	&\b 0		\cr
s			&\b 0											&\t{diag}\left(\left\{e_1 \sqrt{p_{k-r-s+1}},\ldots,e_s \sqrt{p_{k-r}}\right\}\right)
																	&\b 0		\cr
m_e-k+r	&\b 0											&\b 0	&\b 0		\cr
}
\label{eq:He_tilde}
\end{equation}
\hrulefill
\end{figure*}

From Eqs. \eqref{eq:Hb_tilde}, \eqref{eq:He_tilde}, we observe that the new system
equations \eqref{eq:bob_parallel}, \eqref{eq:eve_parallel}
in fact transform the MIMOME system \eqref{eq:bob}, \eqref{eq:eve} into a bank of parallel channels.
Fig. \ref{fig:mimome_to_parallel} shows the resulting parallel channel model. In this parallel channel
model, input symbols $s_1,\ldots,s_{k-r-s}$ are only observed by Eve, symbols $s_{k-r-s+1},\ldots,s_{k-r}$ are received by
both Bob and Eve, whereas symbols $s_{k-r+1},\ldots,s_k$ are only received by Bob.
Finally, symbols $s_{k+1},\ldots,s_{m_a}$ are lost by receivers of both Bob and Eve.

%%%%%%%%%%%%%%%%%%%%%%%%%%%%%%%%%%%%%%%%%%%%%%%%%%%%%%%%%

%%%%%%%%%%%%%%%%%%%%%%%%%%%%%%%%%%%%%%%%%%%%%%%%%%%%%%%%%
\subsection{Reformulation of Secrecy Rate Problem}
\label{sec:secrecy_rate_problem_reformulation}
In this section, we relax the secrecy capacity problem of Eq. \eqref{eq:main_problem}
using the precoding matrix presented in Sec. \ref{sec:precoding_strategy}.
We generalize the problem into an achievable secrecy rate problem for arbitrary input distribution.
To this end, we present the following proposition.

\begin{proposition}
\label{thm:secrecy}
Define $\mathcal{I}\left(\gamma\right) = I\left(s;\sqrt{\gamma}\, s + n\right)$, $\nu = k-r-s$, $\b p = \{p_1,\ldots,p_{m_a}\}$, $\{\omega_i\} = \t{diag}\left(\Omega^H \Omega\right)$. When input $\b s$ of a MIMOME system is a random vector with zero mean entries and identity correlation matrix, by using the precoding matrix $\b W$ defined in Eq. \eqref{eq:precoding_matrix}, we can achieve the following secrecy rate for an arbitrary distribution of $\b s$

\begin{equation}
\begin{split}
&\underset{\b p}{\t{maximize}}~\displaystyle\sum_{i:b_i>e_i} \left[\mathcal{I}\left( \dfrac{b_i ^2}{\omega_{\nu+i}} \, p_{\nu+i}\right) - \mathcal{I}\left( \dfrac{e_i^2}{\omega_{\nu+i}} \, p_{\nu+i}\right)\right]
+\displaystyle\sum_{j=k-r+1}^{k} \mathcal{I}\left(\dfrac{1}{\omega_j}p_j\right)\\
&\t{subject\, to\,:}~~~~
\displaystyle\sum_{i:b_i>e_i} p_{\nu+i} + \displaystyle\sum_{j=k-r+1}^{k} p_j \leq P_T
\end{split}
\label{eq:optimization_parallel}
\end{equation}
\end{proposition}

\begin{proof}
See Appendix \ref{app:theorem}.
\end{proof}

\noindent We observe that, the application of the proposed precoding matrix can transform the MIMOME problem into a distributed 
secrecy rate problem for a bank of parallel channels. We note that, for a given alphabet set, in general 
the above optimization problem should be jointly optimized over the input probability 
distribution and the power allocation. However, practical modulation constellations are 
generally constrained to be equi-probable. Therefore, here we will consider equi-probable input 
alphabet and focus on power allocation optimization. Next we propose a power allocation algorithm to 
solve the above optimization problem.
%%%%%%%%%%%%%%%%%%%%%%%%%%%%%%%%%%%%%%%%%%%%%%%%%%%%%%%%%

%%%%%%%%%%%%%%%%%%%%%%%%%%%%%%%%%%%%%%%%%%%%%%%%%%%%%%%%%
\section{Power allocation algorithm for arbitrary input distribution}
\label{sec:power_allocation}

Even though the above optimization problem is convex for Gaussian input, for an arbitrary input distribution, this is in general not the case. In addition, the lack of closed-form expression for mutual information makes the problem even more difficult to solve. In order to find an efficient sub-optimal solution, we will revert to the decomposition technique. We note that, without the sum power constraint, the optimization problem
\eqref{eq:optimization_parallel} can be decoupled into a number of parallel problems each involving
only one variable $p_i$. However,  the sum power constraint compels us to solve a larger optimization
problem jointly involving multiple variables. For problem such as \eqref{eq:optimization_parallel} involving a
complex constraint, a dual decomposition method \cite{bertsekas1999nonlinear,palomar2006tutorial}
based on the Lagrangian of the objective function enables us to readily decompose the problem
into a number of parallel sub-problems each involving a single variable. These subproblems
are linked through a master problem that updates the dual variable during each
iterations of the subproblems. By introducing a dual variable $\mu$, we can write the
Lagrangian of the optimization problem \eqref{eq:optimization_parallel}
by relaxing the coupling constraint as follows

\begin{equation}
\begin{split}
&\underset{\b p}{\t{maximize}}~
\displaystyle\sum_{i:b_i>e_i} \left[\mathcal{I}\left( \dfrac{b_i ^2}{\omega_{\nu+i}} \, p_{\nu+i}\right) - \mathcal{I}\left( \dfrac{e_i^2}{\omega_{\nu+i}} \, p_{\nu+i}\right)\right]
+\displaystyle\sum_{j=k-r+1}^{k} \mathcal{I}\left(\dfrac{1}{\omega_j}p_j\right)\\
&\quad - \mu\left(\displaystyle\sum_{i:b_i>e_i} p_{\nu+i} + \displaystyle\sum_{j=k-r+1}^{k} p_j\right) + \mu P_T
\\
&\t{subject\, to\,:}~~~~
p_i \geq 0.
\end{split}
%\label{eq:optimization_parallel}
\end{equation}

Note that, the above optimization problem is decoupled in terms of $p_i$.
We obtain the following two subproblems. 

\noindent \textbf{Subproblem 1 :} for all $i$ such that $b_i>e_i$
\begin{equation}
\underset{p_i\geq 0}{\t{maximize}}~ \left[\mathcal{I}\left( \dfrac{b_i ^2}{\omega_{\nu + i}} \, p_{\nu+i}\right) - \mathcal{I}\left(\dfrac{e_i ^2}{\omega_{\nu + i}} \, p_{\nu+i}\right)\right]
- \mu p_{\nu+i}
\end{equation}

\noindent\textbf{Subproblem 2 :} for all $j=k-r+1,\ldots,k$
\begin{equation}
\underset{p_i\geq 0}{\t{maximize}}~ \mathcal{I}\left(\dfrac{1}{\omega_j}\,p_j\right) -  \mu p_j
\end{equation}

\noindent Subproblems 1 and 2 are linked through the master problem which is the dual optimization
problem of \eqref{eq:optimization_parallel}. The master problem updates the value of the dual
variables $\mu$. Let $p_i^*$ denote the solution found from the subproblems.
The master problem can be written as follows

\noindent \textbf{Master Problem :}
\begin{equation}
\begin{split}
&\underset{\mu\geq 0}{\t{minimize}} \displaystyle\sum_{i:b_i>e_i} \left[\mathcal{I}\left(\dfrac{b_i ^2}{\omega_{\nu + i}} \, p_{\nu+i}^*\right) - \mathcal{I}\left(\dfrac{e_i ^2}{\omega_{\nu+i}}\, p_{\nu+i}^*\right)\right]
+\displaystyle\sum_{j=k-r+1}^{k} \mathcal{I}\left(\dfrac{1}{\omega_j}\,p_j^*\right)\\
&\quad - \mu\left(\displaystyle\sum_{i:b_i>e_i} p_{\nu+i}^* + \displaystyle\sum_{j=k-r+1}^{k} p_j^*\right) + \mu P_T .
\end{split}
\label{eq:master_problem}
\end{equation}

Subproblem 2 is a convex problem and can be solve optimally. However, in general subproblem 1 is not convex. Even though subproblems 1 and 2 only involve single variable, in most situations, a closed form or analytic expression for the mutual information is not known for an arbitrary input distribution. In order to solve subproblems 1 and 2,
we resort to a recent result on finite-alphabet research \cite{guo2005mutual,palomar2005gradient}
that relates the mutual information and the minimum mean square error (MMSE) at the receiver
through

\begin{equation}
\dfrac{d\mathcal{I}\left(\rho\right)}{d\rho}=\t{mmse}\left(\rho\right).
\label{eq:mu_mmse}
\end{equation}

\noindent Note that the  function ``mmse'' for different discrete constellations (e.g., $M$-PSK, $M$-QAM etc.,
where $M$ is the number of constellation points) has been given in \cite{lozano2006optimum}.
Using Eq. \eqref{eq:mu_mmse}, the optimum value of $p_i$ in subproblems 2 can be
solved from the following equations
\begin{equation}
\dfrac{1}{\omega_j}\,\t{mmse}\left(\dfrac{1}{\omega_j} p_j^*\right) -  \mu = 0,~~\t{for}~j=k-r+1,\ldots,k.
\label{eq:subproblem2_sol}
\end{equation}
\noindent Eq. \eqref{eq:subproblem2_sol} can be further expressed as follows
\begin{equation}
p_{j}^*= \omega_j \, \t{mmse}^{-1}\left(\min\left\{1,\mu\,\omega_{j}\right\}\right),~~\t{for}~j=k-r+1,\ldots,k
\label{eq:subproblem2_sol_1}
\end{equation}
where we used the fact that $\t{mmse}^{-1}(1) = 0$.

For subproblem 1, we can derive the following necessary condition for optimality
\begin{eqnarray}
\dfrac{b_i^2}{\omega_{\nu+i}} \, \t{mmse}\left(\dfrac{b_i^2}{\omega_{\nu+i}} \, p_{\nu+i}^*\right)
- \dfrac{e_i^2}{\omega_{\nu+i}}\, \t{mmse}\left(\dfrac{e_i^2}{\omega_{\nu+i}} \, p_{\nu+i}^*\right)
- \mu = 0,\nonumber\\
\qquad\qquad\qquad\qquad\qquad\qquad\qquad\qquad\t{for}~i:b_i>e_i\qquad
\label{eq:subproblem1_sol}
\end{eqnarray}

Next, we propose a sufficient condition for the optimality of the solution of Eq. \eqref{eq:subproblem1_sol}. In this regard, we define the following MMSE difference function
\begin{equation}
f_{\t{mmseD}} (p,b_i,e_i,\omega_{\nu+i}) = \dfrac{b_i^2}{\omega_{\nu+i}} \, \t{mmse}\left(\dfrac{b_i^2}{\omega_{\nu+i}} \, p_{\nu+i}^*\right)
- \dfrac{e_i^2}{\omega_{\nu+i}}\, \t{mmse}\left(\dfrac{e_i^2}{\omega_{\nu+i}} \, p_{\nu+i}^*\right).
\label{eq:mmseD}
\end{equation}
We note that, a similar condition has been proposed in \cite{rodrigues2010gaussian} [see Theorem 6 in \cite{rodrigues2010gaussian}] for a parallel Gaussian wiretap channel with $M$-PAM inputs.

\begin{proposition}
\label{prop:sufficient_condition}
If the MMSE difference function in Eq. \eqref{eq:mmseD} admits a unique zero $p'$ and is strictly monotonically decreasing for $0\leq p\leq p'$, then the optimal solution $\left\{ p_{\nu+i}^{*} \right\}$ of Eq. \eqref{eq:subproblem1_sol} can be given as follows

when \textbf{$b_{i}^{2}\leq e_{i}^{2}$}

$\quad\quad\quad\quad p_{\nu+i}^{*}=0$

when \textbf{$b_{i}^{2}>e_{i}^{2}$}

$\quad\quad\quad\quad p_{\nu+i}^{*}=p'\quad\t{if\,}\mu=0$

$\quad\quad\quad\quad \dfrac{b_i^2}{\omega_{\nu+i}} \, \t{mmse}\left(\dfrac{b_i^2}{\omega_{\nu+i}} \, p_{\nu+i}^*\right) - \dfrac{e_i^2}{\omega_{\nu+i}}\, \t{mmse}\left(\dfrac{e_i^2}{\omega_{\nu+i}} \, p_{\nu+i}^*\right) = \mu, \quad0<p<p'\quad\t{if\,}\mu>0$.
\end{proposition}

\begin{proof}
See Appendix \ref{app:sufficient_condition}.
\end{proof}
In following proposition, we describe a condition for the MMSE difference function to admit a unique zero solution.

\begin{proposition}
\label{prop:unique_zero}
If $g(\rho) = \rho\, \t{mmse}(\rho)$ is a strictly uni-modal function and  $b_i > e_i$, then the MMSE difference function $f_{\t{mmseD}} (p,b_i,e_i,\omega_{\nu + i})$ admits a unique zero solution for $p >0$.
\end{proposition}

\begin{proof}
See Appendix \ref{app:unique_zero}.
\end{proof}

\begin{figure}[ht]
\centering
\subfigure[]{
\includegraphics[scale=0.8]{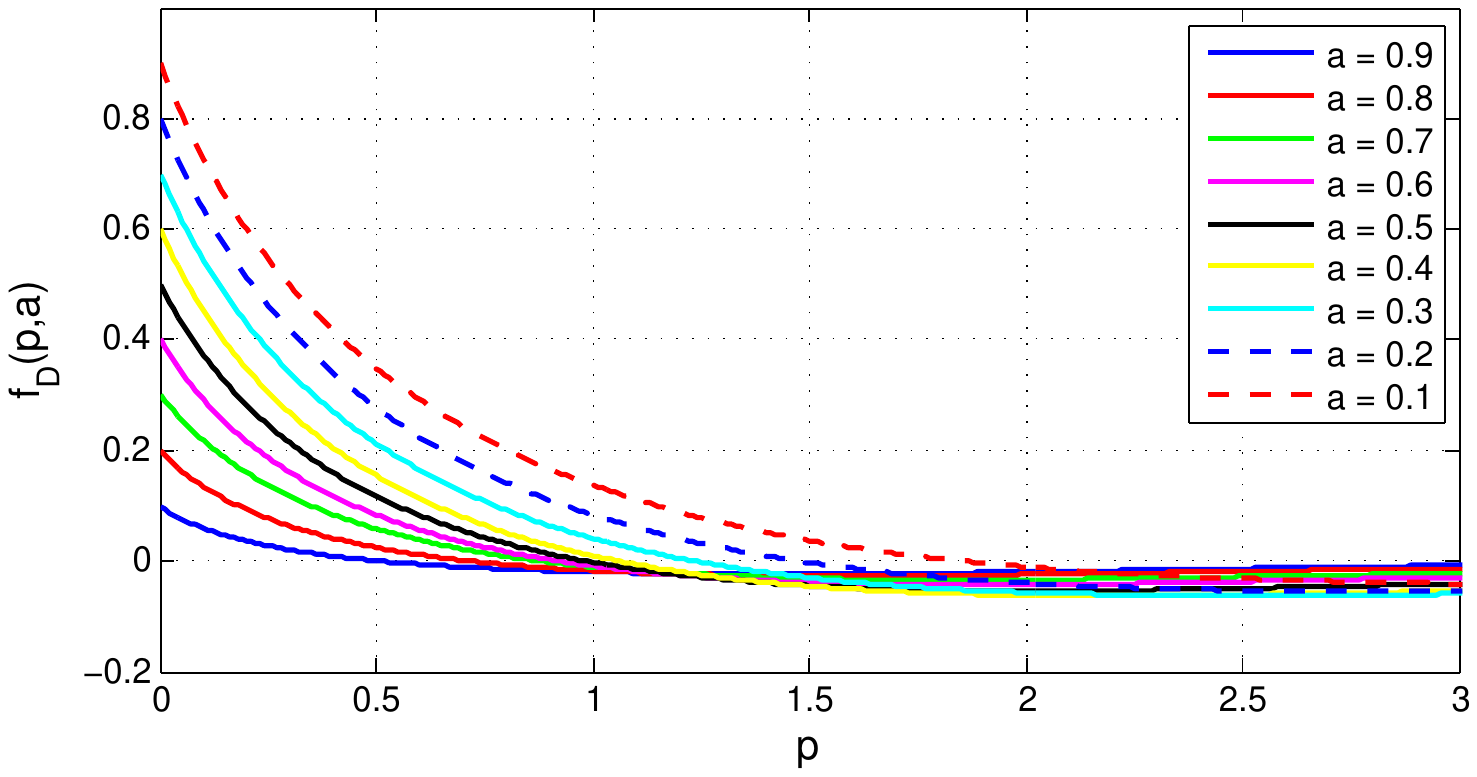}
\label{fig:subfig1_bpsk}
}
\subfigure[]{
\includegraphics[scale=0.8]{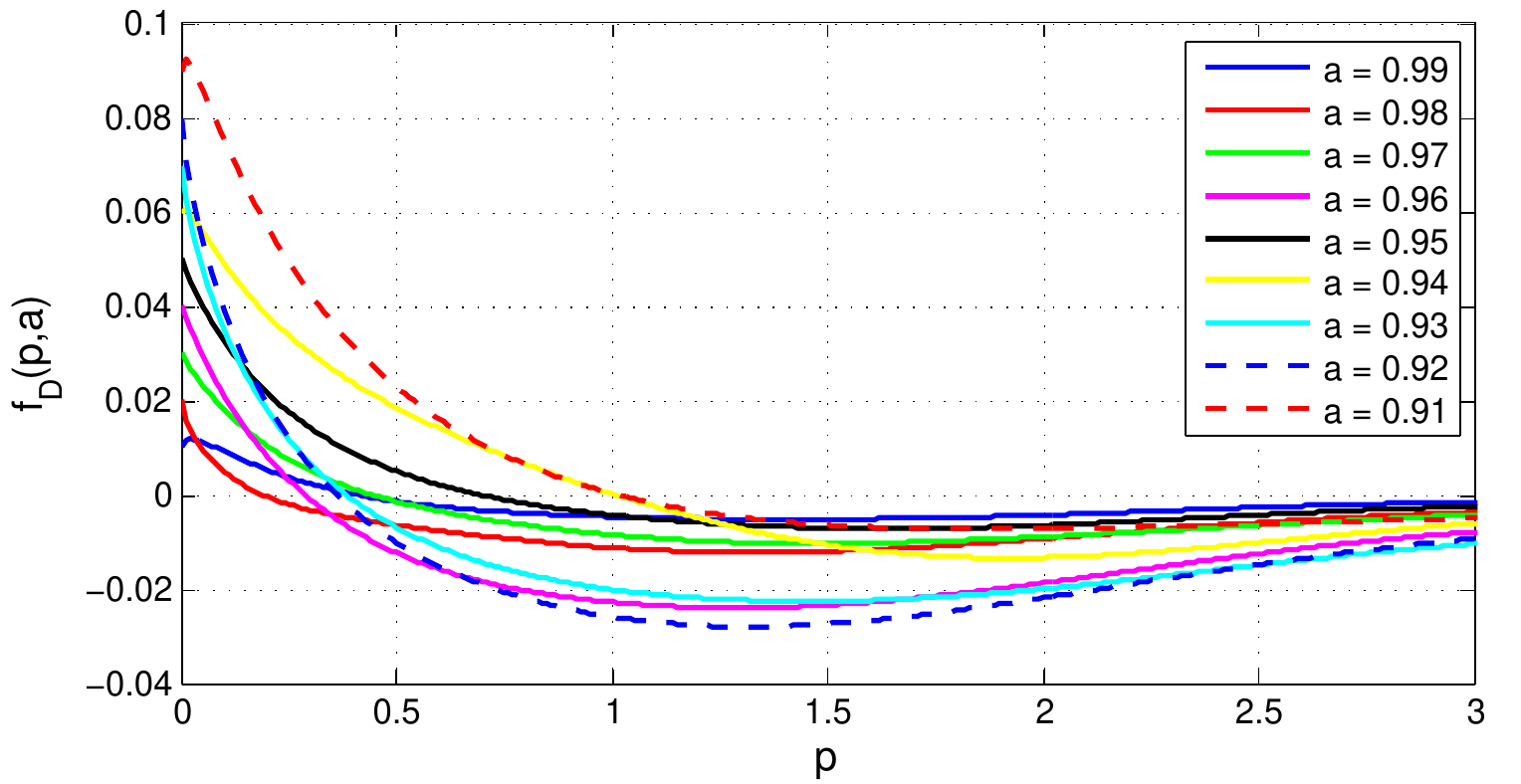}
\label{fig:subfig2_bpsk}
}
\subfigure[]{
\includegraphics[scale=0.8]{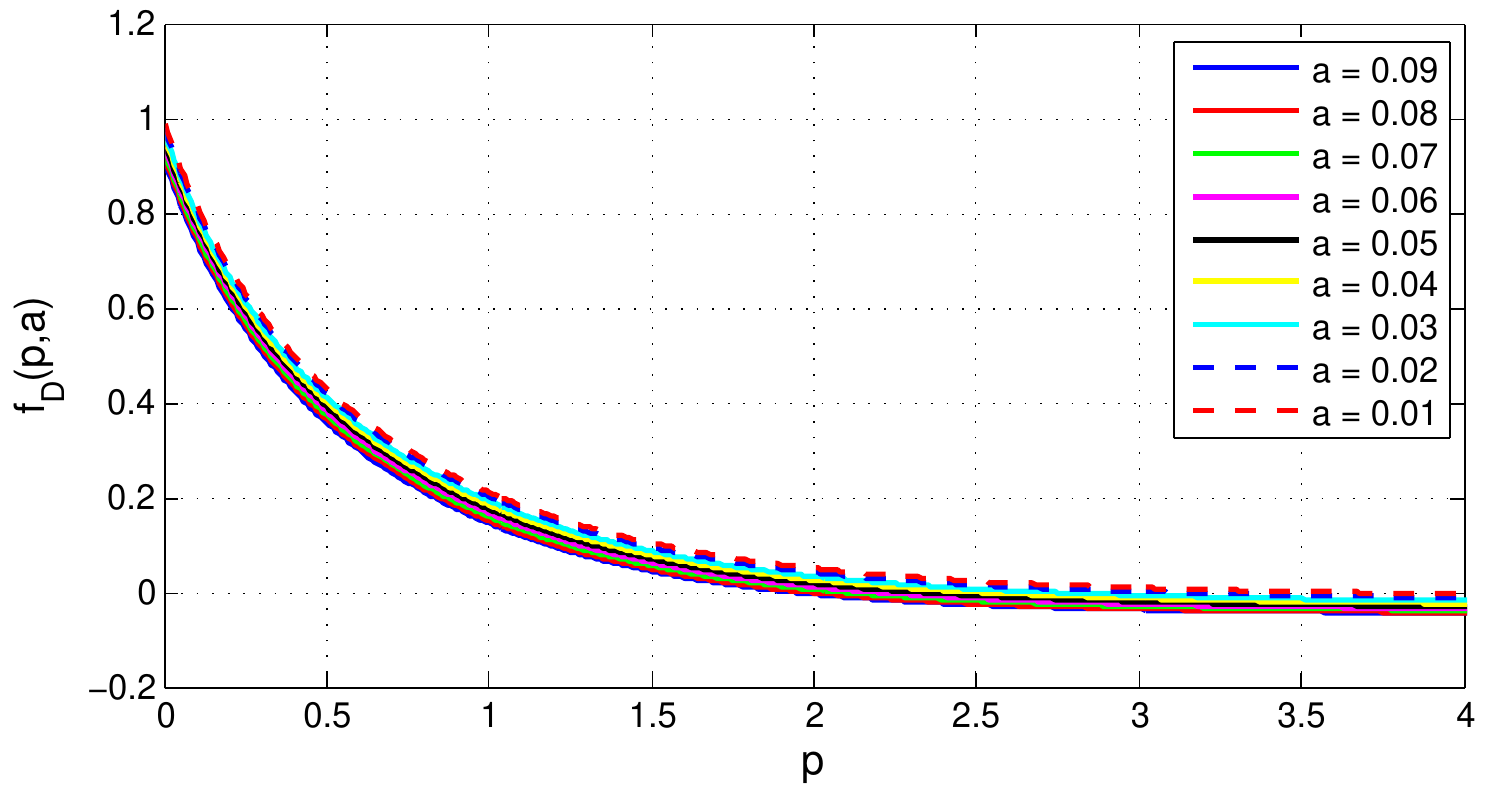}
\label{fig:subfig3_bpsk}
}
\caption[]{MMSE difference function $f_D(p,a)$ for different values of $a = e_i^2 \diagup b_i^2$ for BPSK}
\label{fig:mmseD_bpsk}
\end{figure}

\begin{figure}[ht]
\centering
\subfigure[]{
\includegraphics[scale=0.8]{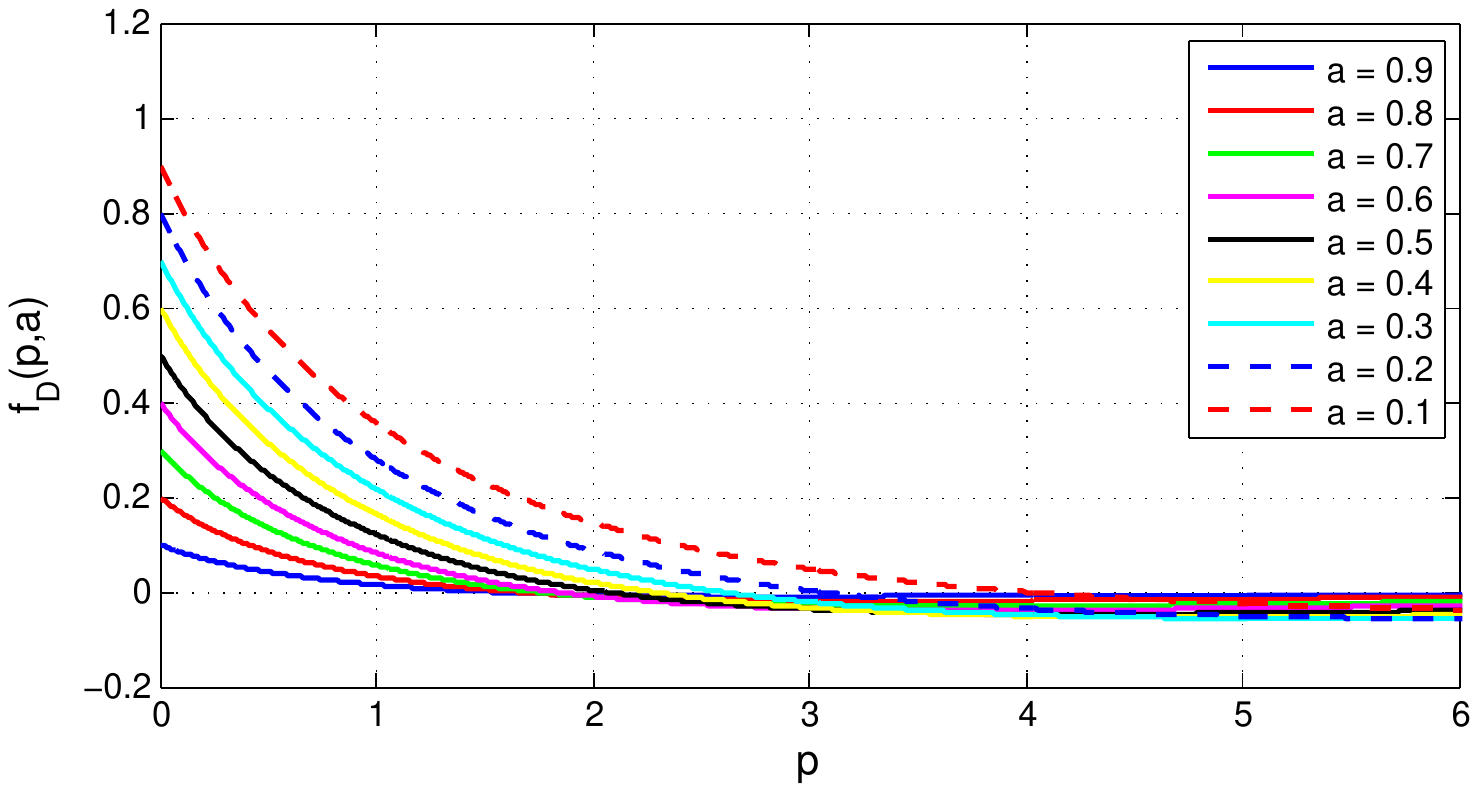}
\label{fig:subfig1_qpsk}
}
\subfigure[]{
\includegraphics[scale=0.8]{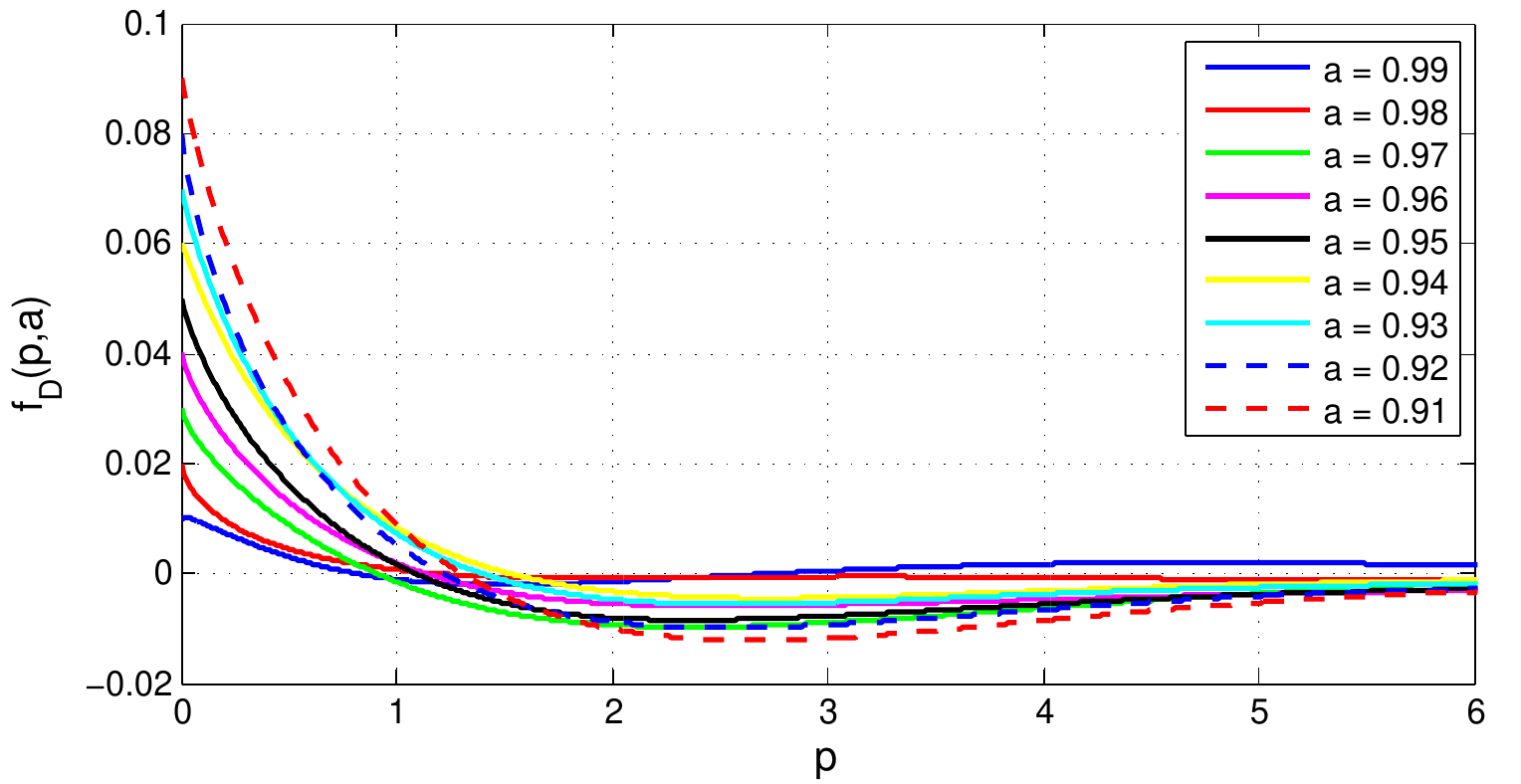}
\label{fig:subfig2_qpsk}
}
\subfigure[]{
\includegraphics[scale=0.8]{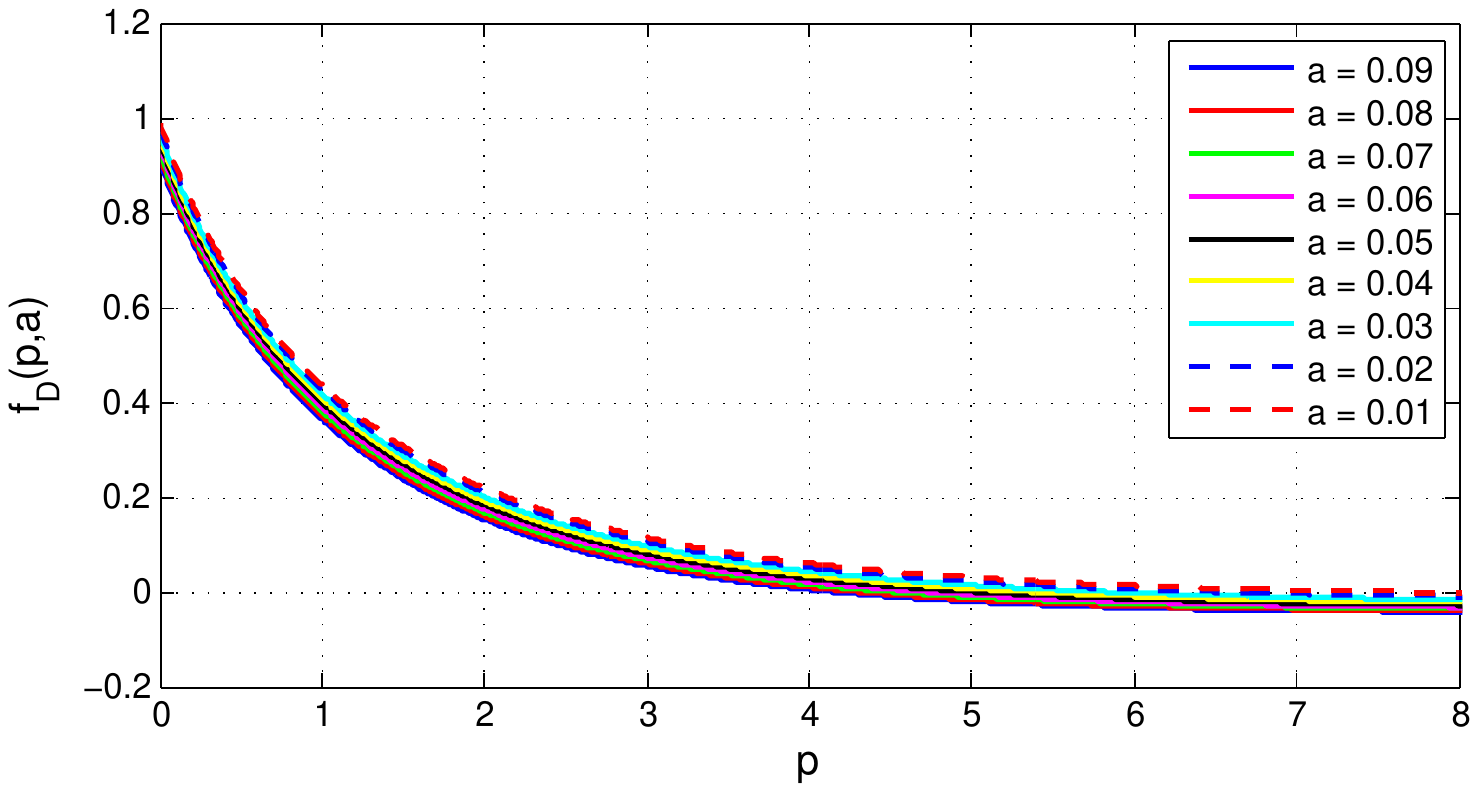}
\label{fig:subfig3_qpsk}
}
\caption[]{MMSE difference function $f_D(p,a)$ for different values of $a = e_i^2 \diagup b_i^2$ for QPSK}
\label{fig:mmseD_qpsk}
\end{figure}

\begin{figure}[ht]
\centering
\subfigure[]{
\includegraphics[scale=0.8]{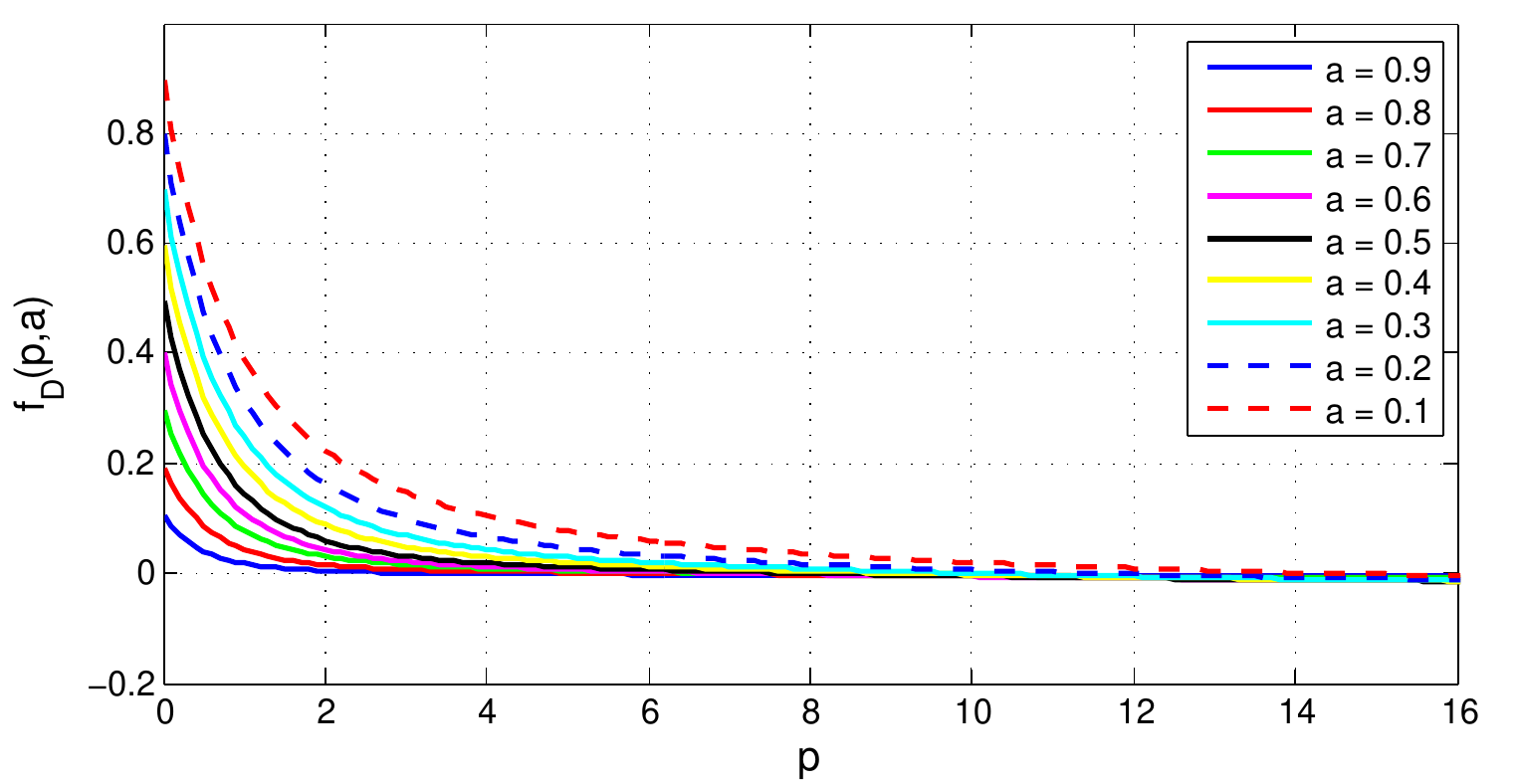}
\label{fig:subfig1_16qam}
}
\subfigure[]{
\includegraphics[scale=0.8]{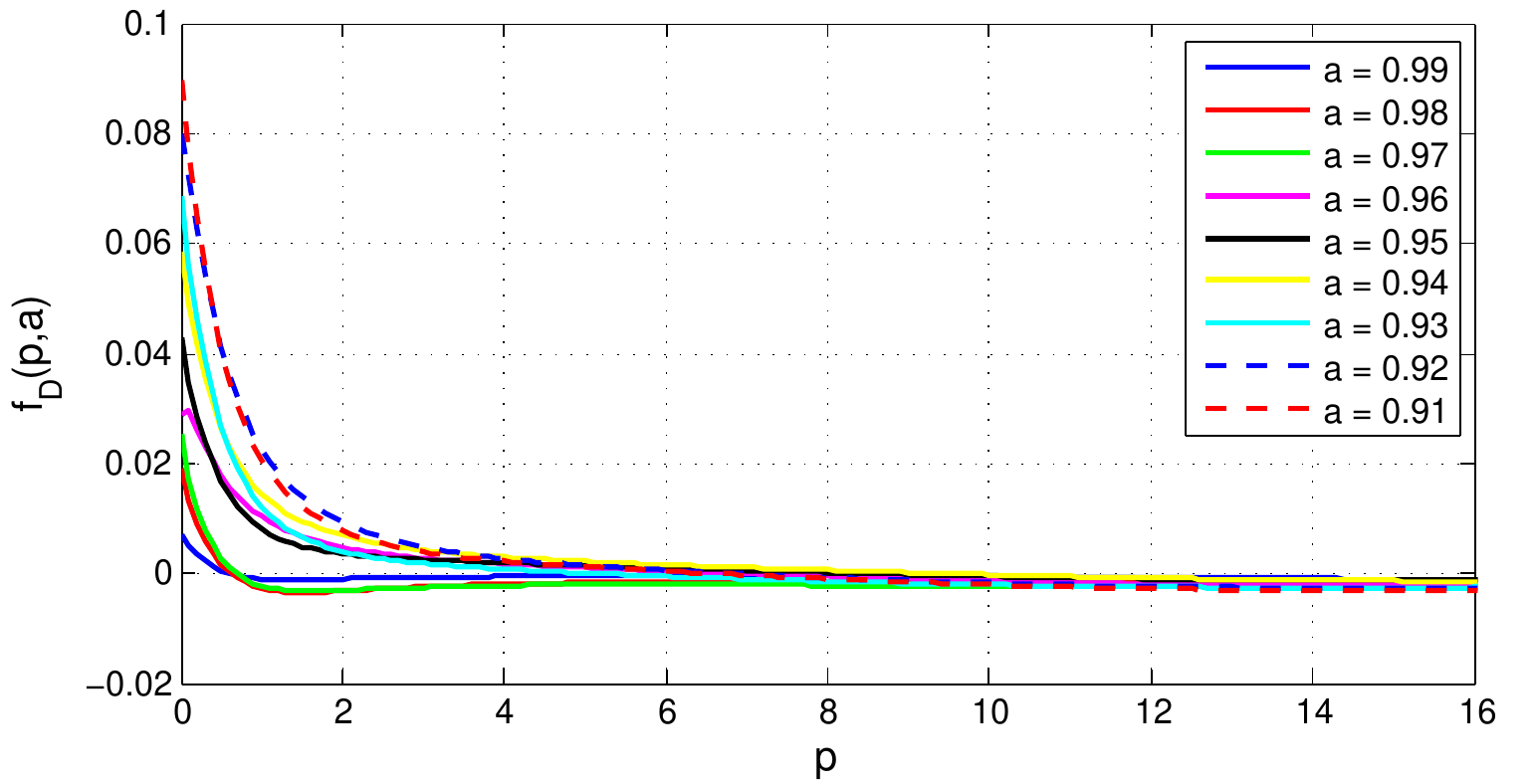}
\label{fig:subfig2_16qam}
}
\subfigure[]{
\includegraphics[scale=0.8]{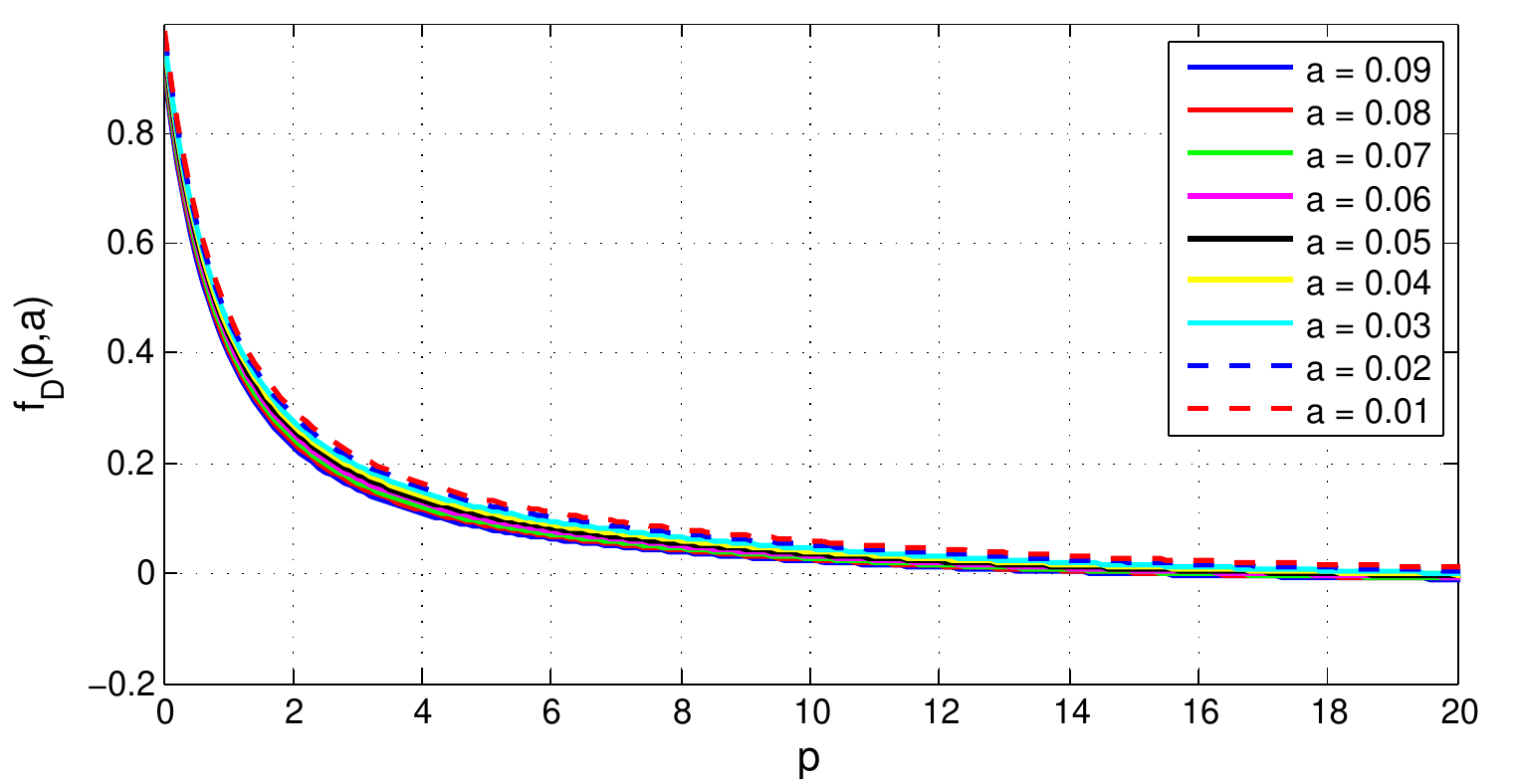}
\label{fig:subfig3_16qam}
}
\caption[]{MMSE difference function $f_D(p,a)$ for different values of $a = e_i^2 \diagup b_i^2$ for $16$-QAM}
\label{fig:mmseD_16qam}
\end{figure}

\begin{figure}[ht]
\centering
\subfigure[]{
\includegraphics[scale=0.8]{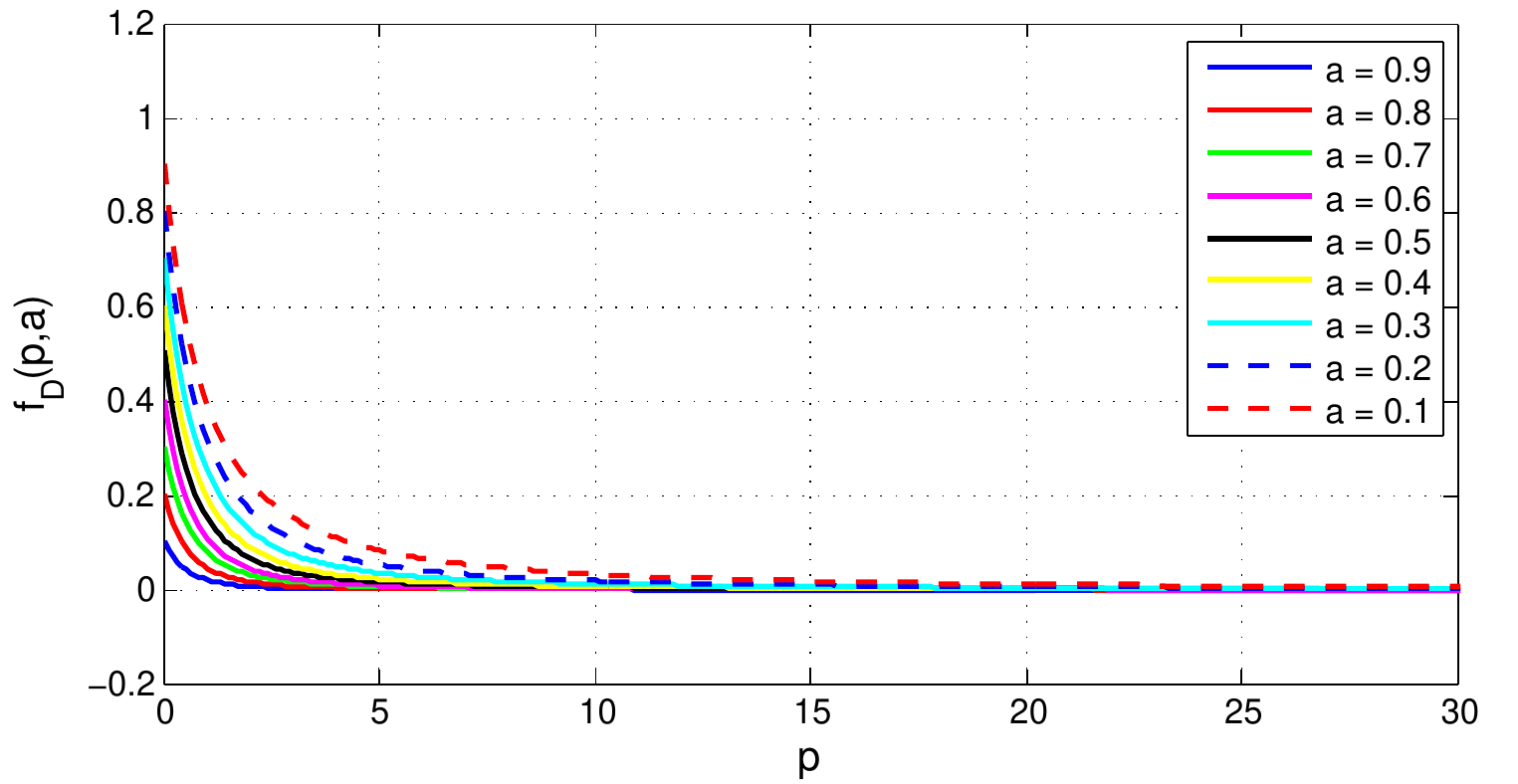}
\label{fig:subfig1_64qam}
}
\subfigure[]{
\includegraphics[scale=0.8]{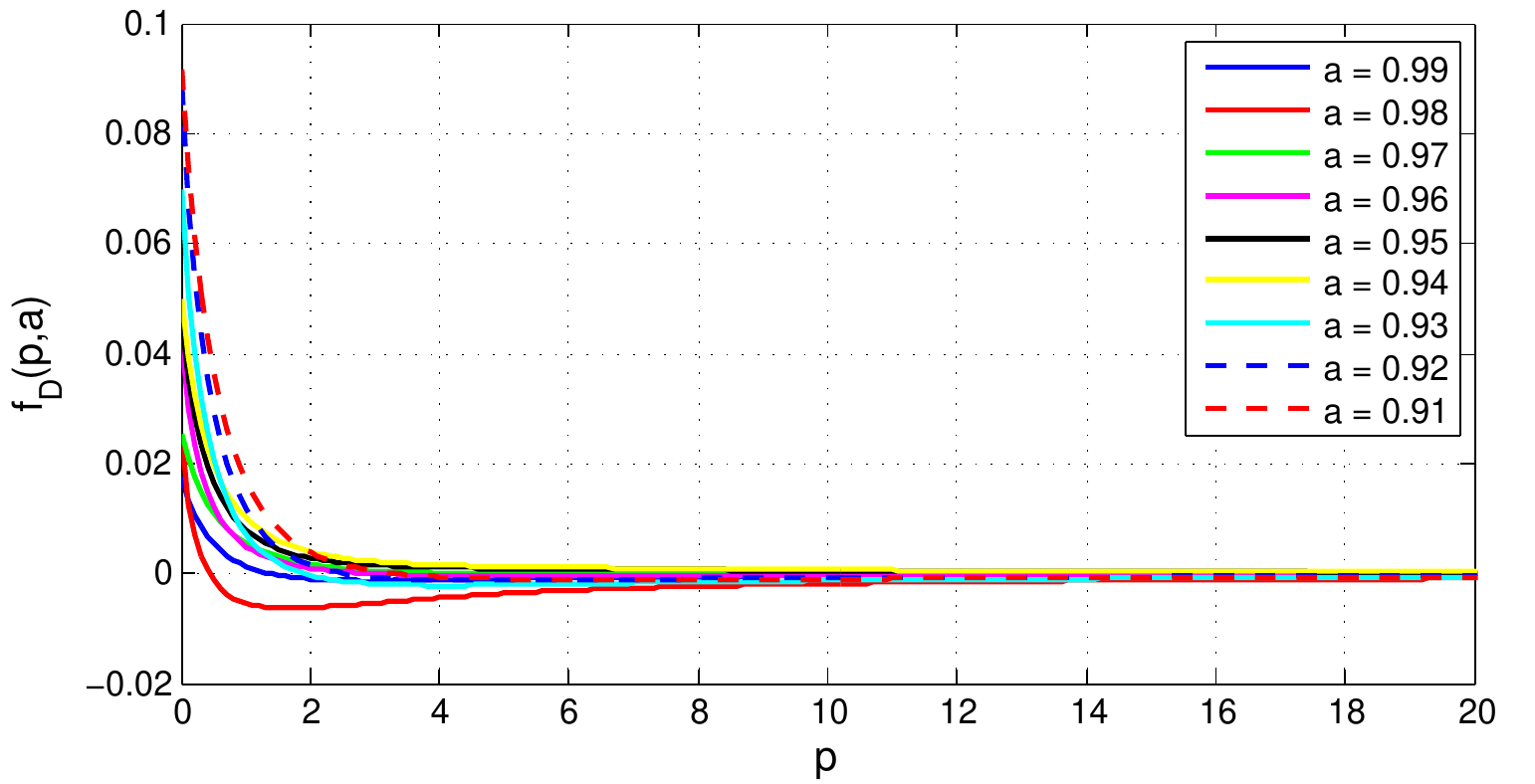}
\label{fig:subfig2_64qam}
}
\subfigure[]{
\includegraphics[scale=0.8]{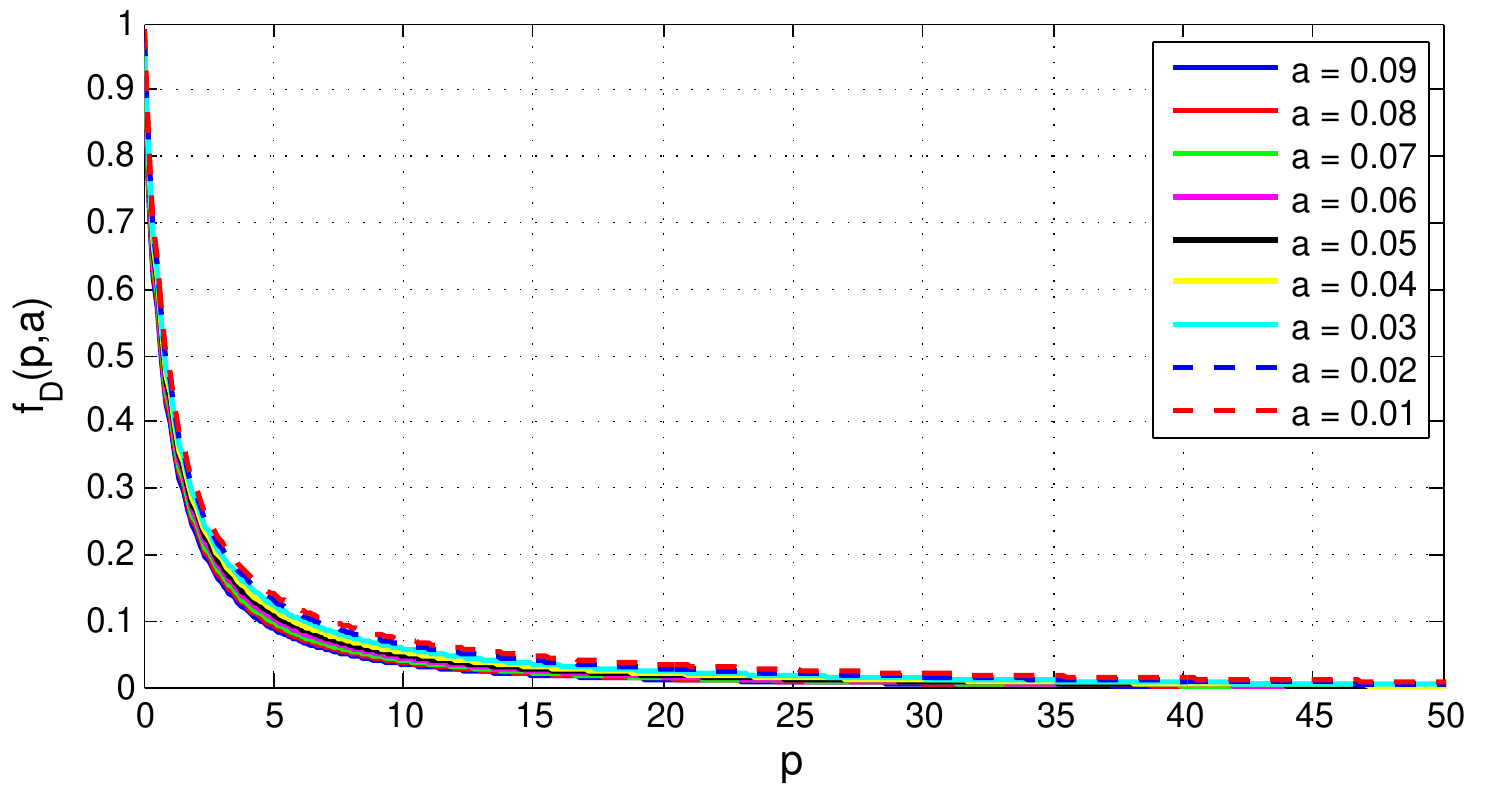}
\label{fig:subfig3_64qam}
}
\caption[]{MMSE difference function $f_D(p,a)$ for different values of $a = e_i^2 \diagup b_i^2$ for $64$-QAM}
\label{fig:mmseD_64qam}
\end{figure}

In Figure \ref{fig:rhommse}, we present the plot of $g(\rho) = \rho\, \t{mmse}(\rho)$ vs. $\rho$ for BPSK, QPSK, $16$-QAM, and $64$-QAM input constellations. We observe that
$g(\rho)$ of all four constellations shows strictly uni-modality in the region of interest. 
Furthermore, in Figures \ref{fig:mmseD_bpsk}, \ref{fig:mmseD_qpsk}, \ref{fig:mmseD_16qam}, and \ref{fig:mmseD_64qam}, we illustrate 
graphically the MMSE difference function $f_{\t{mmseD}} (p,a)$ vs. power $p$ for various values of $a = e_i^2 \diagup b_i^2$. Here, we also assume $\omega_{\nu + i} =1$ without any loss of generality. We observe that, the MMSE difference function exhibits 
strictly monotonically decreasing behavior in the range $0\leq p\leq p'$ for all four constellations. Based on 
Propositions \ref{prop:sufficient_condition} and \ref{prop:unique_zero}, and Figures \ref{fig:rhommse}-\ref{fig:mmseD_64qam}, 
we see that the optimality criterion described in proposition \ref{prop:sufficient_condition} holds for common constellations of BPSK, 
QPSK, $16$-QAM, and $64$-QAM. Therefore, the solution of subproblem 1 in Eq. \eqref{eq:subproblem2_sol_1} 
will be unique for at least these four constellations.

The combined algorithm to solve the optimization problem \eqref{eq:optimization_parallel} is presented in Algorithm \eqref{alg:dual_decomp}.

\begin{algorithm}[H]
\caption{Dual decomposition algorithm for \eqref{eq:optimization_parallel}}
\label{alg:dual_decomp}
\begin{enumerate}
\item Initialize dual variable $\mu\geq 0$.
\item Find solution of subproblem 1 and 2 by solving  \eqref{eq:subproblem1_sol} and \eqref{eq:subproblem2_sol_1}, respectively.
\item Update dual variable as\\
\hspace*{-.5cm}$\mu = \left[\mu + \alpha\left(\displaystyle\sum_{i:b_i>e_i} p_{\nu+i} + \displaystyle\sum_{j=k-r+1}^{k} p_j - P_T \right)\right]^+$
\item Go to Step 2 until stopping criterion is reached.
\end{enumerate}
\end{algorithm}

We note that the master problem in Eq. \eqref{eq:master_problem} is differentiable with respect to the dual variable $\mu$. Therefore, in Step 3 of the above algorithm, we use the gradient method to update the dual variable. Here, the parameter $\alpha > 0$ denotes an appropriate stepsize, which can be 
either a constant or time-varying. In our simulation, we choose a fixed stepsize $\alpha$. If the stepsize 
is sufficiently small, then the solution of the above algorithm will converge to the solution of the optimal dual variable $\mu^*$. 
A detail description on the choice of stepsize and stopping criterion can be found in \cite{bertsekas1999nonlinear,palomar2006tutorial} 
and the references therein. If the original optimization problem in Eq. \eqref{eq:optimization_parallel} is convex, then the duality gap will be zero. Therefore, 
the solution of the dual problem in Eq. \eqref{eq:master_problem} will also provide the optimal solution of the original problem. 
However, if the original problem is not convex, then there exists a positive duality gap and the solution of Eq. \eqref{eq:master_problem} will 
be a suboptimal solution of the original problem.

%%%%%%%%%%%%%%%%%%%%%%%%%%%%%%%%%%%%%%%%%%%%%%%%%%%%%%%%%%
\section{Secrecy Rate Analysis of MIMOME System}
\label{sec:analysis}

\subsection{Gaussian Input Case}
\label{sec:gaussian}

In this section, we will present the power allocation problem for the special case of Gaussian input distribution based 
on the framework considered above. Even though
the result for Gaussian input is well known \cite{li2010secrecy, khisti_MIMOME, liang2008secure}, 
results in this section will provide additional insight for the finite-alphabet input scenarios to be considered later.

For Gaussian input, the MMSE equation is simply
\begin{equation}
\t{mmse}(\gamma) = \dfrac{1}{1+\gamma}.
\label{eq:mmse_gaussian}
\end{equation}
Let us denote $\left\{ p_i^{\t{g}} \right\}$ as the optimum power allocation for Gaussian input. 
Based on Eq. \eqref{eq:mmse_gaussian}, solutions of subproblem 1 can be found by solving the following equation
\begin{equation}
\dfrac{b_i^2 e_i^2}{b_i^2 - e_i^2}\dfrac{1}{\omega_{\nu + 1}}\, \left(p_{\nu + 1}^{\t{g}}\right)^2 + \dfrac{1}{b_i^2 - e_i^2} p_{\nu + 1}^{\t{g}} +\left[\dfrac{\omega_{\nu + 1}}{b_i^2 - e_i^2} -\dfrac{1}{\mu} \right] = 0.
\label{eq:soln_gaussian}
\end{equation}
Thus, the optimal power allocation is 
\begin{equation}
p_{\nu+i}^{\t{g}} = \begin{cases}
0, & \t{if\,}\dfrac{1}{\mu} \leq \dfrac{\omega_{\nu+i}}{b_i^2 - e_i^2}\\
\dfrac{1}{2}\left[\sqrt{\left(\dfrac{\omega_{\nu+i}}{b_{i}^{2}e_{i}^{2}}\right)^{2}+4\dfrac{\omega_{\nu+i}}{b_{i}^{2}e_{i}^{2}}\left(b_{i}^{2}-e_{i}^{2}\right)\left(\dfrac{1}{\mu}-\dfrac{\omega_{\nu+i}}{b_{i}^{2}-e_{i}^{2}}\right)}-\dfrac{\omega_{\nu+i}}{b_{i}^{2}e_{i}^{2}}\right], & \t{if\,}\dfrac{1}{\mu}>\dfrac{\omega_{\nu+i}}{b_i^2 -e_i^2}\end{cases}.
\label{eq:subproblem1_gauss}
\end{equation}
Similarly, based on Eq. \eqref{eq:mmse_gaussian}, solution of subproblem 2 can be found as
\begin{equation}
p_{i}^{\t{g}}=\begin{cases}
0, & \t{if}\,\dfrac{1}{\mu}\leq \omega_{i}\\
\dfrac{1}{\mu}-\omega_i, & \t{if}\,\dfrac{1}{\mu}>\omega_{i}\end{cases} .
\label{eq:subproblem2_gauss}
\end{equation}

We notice that, the solution of subproblem 2 presented in Eq. \eqref{eq:subproblem2_gauss} can also be obtained 
from Eq. \eqref{eq:soln_gaussian} for by replacing $b_i^2$ with $1$ and $e_i^2$ with $0$. In the absence of eavesdropper, the solution of this simplified 
problem is the famous water-filling solution as given in Eq. \eqref{eq:subproblem2_gauss}, where the 
base level is $\omega_i$ and water level is the inverse of the dual variable $\dfrac{1}{\mu}$. 
When the security constraint is present, however, the problem takes on an interesting structure. 
In such case, we can still consider the water level as $\dfrac{1}{\mu}$. However, we will use 
$\dfrac{\omega_{\nu+i}}{b_i^2 -e_i^2}$ as a base level to take into account the additional constraint due to secrecy. 
For the parallel channels $j=k-r+1,\ldots,k$, with components only towards Bob's direction, 
the base level will still be $\omega_j$ (since, $b_{j-\nu}^2 = 1$ and $e_{j-\nu}^2 = 0$). Similar to water-filling, 
we will allocate power only when the water level is above the base level. 
However, the power level in this case will not be the difference between the water level and base level,
 i.e. $\dfrac{1}{\mu} - \dfrac{\omega_{\nu+i}}{b_i^2 -e_i^2}$.
Instead, it will be a non-linear function of the difference as given in Eq. \eqref{eq:subproblem1_gauss}. 
As a result, the achievable secrecy rate can be written as

\begin{equation}
R_{s}^{\t{g}} = \displaystyle\sum_{i:b_i>e_i} \log\left(\dfrac{1+\dfrac{b_i^2}{\omega_{\nu+i}}\,p_{\nu+i}^{\t{g}}}{1+\dfrac{e_i^2}{\omega_{\nu+i}}\,p_{\nu+i}^{\t{g}}}\right)
+\displaystyle\sum_{j=k-r+1}^{k} \log\left(1+\dfrac{1}{\omega_j}\,p_j^{\t{g}}\right) .
\label{eq:rate_gauss}
\end{equation}

%We show that in the high SNR regime, the proposed precoding strategy achieves the
%upper bound presented in \cite{khisti_MIMOME}. For Gaussian input, the mmse equation becomes
%\begin{equation}
%\t{mmse}(\gamma) = \dfrac{1}{1+\gamma}.
%\label{eq:mmse_gaussian}
%\end{equation}
%
%
%Denote $\sigma_i = \dfrac{b_i}{e_i}$.
%In the high SNR regime, we can approximate Eq. \eqref{eq:rate_gauss} as
%\begin{equation}
%R_{s}^{\t{g}} = \displaystyle\sum_{i:\sigma_i > 1} \log\sigma_i^2 +\displaystyle\sum_{j=k-r+1}^{k} \log\left(p_j^{\t{g}}\right) .
%\label{eq:rate_gauss_highsnr}
%\end{equation}
%Note that, if $\t{rank}\left(\b H_e\right)=m_a$,  $\t{dim}\left(\mathcal{S}_b\right)=r=0$. Then the high
%SNR approximation in Eq. \eqref{eq:rate_gauss_highsnr} becomes the same high SNR approximation provided in \cite{khisti_MIMOME}. However, if
%$\t{rank}\left(\b H_e\right)<m_a$, the high SNR approximation (Eq. \eqref{eq:rate_gauss_highsnr}) differs
%from the high SNR approximation in \cite{khisti_MIMOME} because of the different precoding structure in use.
%%%%%%%%%%%%%%%%%%%%%%%%%%%%%%%%%%%%%%%%%%%%%%%%%%%%%%%%%%

%%%%%%%%%%%%%%%%%%%%%%%%%%%%%%%%%%%%%%%%%%%%%%%%%%%%%%%%%
\subsection{Low SNR Approximation}
\label{sec:low_snr}

\subsubsection{Second order optimal signaling}
For second order optimal signaling \cite{verdu2002spectral}, the first
and the second order derivatives of the mutual information achieved at zero SNR
matches with those achieved using Gaussian input. In general, quadratic symmetric signaling 
such as QPSK or any other signaling distribution that can be written as a mixture of 
QPSK (i.e., $M$-QAM, for $M\geq 4$) are second order optimal. For second order optimal signaling,
the low SNR approximation of MMSE (i.e., the first derivative of mutual information)
is the same as that of Gaussian signaling. Hence, we can use the same water-filling power allocation solution as presented in Eq. \eqref{eq:subproblem1_gauss}, \eqref{eq:subproblem2_gauss} in Section \ref{sec:gaussian}.

\subsubsection{Non-second order optimal signaling}
$1$-D signaling schemes such as BPSK and $M$-PAM are not
second order optimal. A low SNR approximation of such signaling is given in \cite{lozano2006optimum} as

\begin{equation}
\t{mmse}\left(\rho\right)=1-2\,\rho+o\left(\rho^{2}\right) .
\end{equation}

Based on the above equation, low SNR power allocation $\left\{p_i^{\t{low}}\right\}$ for non-second order optimal signaling are given below in two cases:

\begin{align}
&p_{\nu+i}^{\t{low}}=\begin{cases}
0, & \t{when\,} \dfrac{1}{\mu} \leq \dfrac{\omega_{\nu+i}}{b_i^2 -e_i^2} \\
\dfrac{\omega_{\nu+i}}{2} \mu \left( \dfrac{1}{\mu} - \dfrac{\omega_{\nu+i}}{b_i^2 - e_i^2}
\right) & \t{when\,}
\dfrac{1}{\mu} > \dfrac{\omega_{\nu+i}}{b_i^2 -e_i^2}
\end{cases},
\quad\{i:\;b_i>e_i\}
\label{eq:palloc_nonsecopt_1}
\end{align}

\begin{align}
&p_{i}^{\t{low}}=\begin{cases}
0, & \t{when}\, \dfrac{1}{\mu} \leq \omega_i \\
\dfrac{\omega_i}{2} \mu \left( \dfrac{1}{\mu} -  \omega_{i} \right), & \t{when}\, \dfrac{1}{\mu} > \omega_i\end{cases}, & i=k-r+1,\ldots,k.
\label{eq:palloc_nonsecopt_2}
\end{align}
The achievable secrecy rate at low SNR can be approximated as follows
\begin{equation}
R_s^{\t{low}}=\displaystyle\sum_{i:b_i>e_i}
\left(b_i^2-e_i^2\right)\left(p_{\nu+i}^{\t{low}}-\left(p_{\nu+i}^{\t{low}}\right)^2\right) +\displaystyle\sum_{j=k-r+1}^{k} \left(p_{j}^{\t{low}}-\left(p_{j}^{\t{low}}\right)^2\right).
\end{equation}

We notice that, similar to the case of Gaussian and second-order optimal signaling, we 
can use a water-filling strategy with water level $\dfrac{1}{\mu}$ and base level 
$\dfrac{\omega_{\nu+i}}{b_i^2 -e_i^2}$ and assuming $b_{j-\nu}^2 = 1$ and $e_{j-\nu}^2 = 0$ for $j=k-r+1,\ldots,k$. The power level is still function of 
\[\dfrac{1}{\mu} - \dfrac{\omega_{\nu+i}}{b_i^2 -e_i^2},\]
which is the difference between water level and base level. 
The functions are now slightly different in (Eq.  \eqref{eq:palloc_nonsecopt_1} and \eqref{eq:palloc_nonsecopt_2}).

%%%%%%%%%%%%%%%%%%%%%%%%%%%%%%%%%%%%%%%%%%%%%%%%%%%%%%%%%%

%%%%%%%%%%%%%%%%%%%%%%%%%%%%%%%%%%%%%%%%%%%%%%%%%%%%%%%%%%
\subsection{High SNR Approximation}
\label{sec:high_snr}

\subsubsection{ $\t{rank} \left( \b H_e \right) = m_a$ }

In this case $\t{dim} \left( \b \mathcal{S}_b \right) = r = 0$.
For the power constraint in Eq. \eqref{eq:optimization_parallel},
we get the following complementary slackness condition

\begin{equation}
\mu \left(\displaystyle\sum_{i:b_i>e_i} \omega_{\nu+i} \, p_{\nu+i} - P_T \right) = 0.
\end{equation}

Based on proposition \ref{prop:sufficient_condition}, the optimal power $p^*$ for maximum secrecy
satisfies $p^* \leq p'$. Therefore, at very high SNR when $P_T\to\infty$, 
secrecy rate for finite-alphabet input is maximized by using a fraction of the total available power. 
The power constraint
inequality becomes a strict inequality. In the above complementary slackness condition,
we attain $\mu=0$. Denoting high SNR power allocation as $p^{\t{high}}_i$,
we re-write Eq. \eqref{eq:subproblem1_sol} as

\begin{equation}
b_i^2\, \t{mmse}\left(b_i^2\, p_{\nu+i}^{\t{high}}\right) = e_i^2\, \t{mmse}\left(e_i^2\, p_{\nu+i}^{\t{high}}\right).
\end{equation}

In \cite{lozano2006optimum}, based on the sub-optimum estimator
$\hat{s}(y,\rho) = \arg\underset{s_k}{\min} \abs{y-\sqrt{\rho}\,s_k}$ the
following MMSE approximation at  high SNR is found:
\begin{equation}
\t{mmse}(\rho) \approx K\,\exp\left\{-\dfrac{d^2}{4}\,\rho\right\},
\label{eq:high_snr_approximation}
\end{equation}
in which $K$ is a constant and $d$ is the minimum distance between two signaling
points in the discrete unit variance input constellation.

Reference \cite{lozano2006optimum} also provided a table containing formula
for calculating $d$ of different finite-alphabet constellations.
Using Eq. \eqref{eq:high_snr_approximation}, we obtain the following high SNR
approximation of power allocation
\begin{equation}
p_{\nu+i}^{\t{high}}=\dfrac{\log\sigma_{i}^{2}}{\dfrac{d^{2}}{4}
\dfrac{\left(b_{i}^{2}-e_{i}^{2}\right)}{\omega_{\nu+i}}},~~\t{for}~i~\t{such~that}~b_i>e_i .
\label{eq:power_alloc_highsnr}
\end{equation}

In \cite{lozano2006optimum}, it was observed that, at high SNR regime, the 
power allocation for parallel Gaussian channel with finite-alphabet demonstrates 
a channel inversion characteristic. In other words, stronger channels receive less power allocation.
This is because the mutual information of a $M$-ary constellation cannot exceed $\log_2 M$ bits/s/Hz,
Thus, there is little incentive to allocate more power to a channel once the mutual information is near saturation.
Instead, additional power is better allocated to weaker channels for higher rate. 
In Eq. \eqref{eq:power_alloc_highsnr}, we observe a similar channel inversion phenomenon,
although in this case, the effective channel $\dfrac{\left(b_{i}^{2}-e_{i}^{2}\right)}{\omega_{\nu+i}}$. 
This is in sharp contrast to both the water-filling power allocation at low SNR regime and the power allocation
for Gaussian input (see Eq. \eqref{eq:palloc_nonsecopt_1}), where 
the power allocation was proportional to the effective channel.

\subsubsection{$\t{rank}\left(\b H_e\right) < m_a$}
Based on Eq. \eqref{eq:high_snr_approximation}, mmse($\rho$) decays exponentially to zero as $\rho \to \infty$. 
From Eq. \eqref{eq:subproblem2_sol}, therefore, 
we find that $\mu\to 0$ as $P_T\to\infty$. Hence, for the subset of parallel channels, 
$i:b_i > e_i$, Eq. \eqref{eq:power_alloc_highsnr} will still provide a high SNR approximation of power allocation.

The subset of parallel channels  $j=k-r+1,\ldots,k$ are in the subspace $\mathcal{S}_b$. 
For these channels, there are no components in Eve's subspace and a similar channel inversion style power allocation can
be achieved as presented in \cite{lozano2006optimum} with an effective channel $\dfrac{1}{\omega_j}$. 
For an $M$-ary constellation, mutual information for these channels will become close to $\log_2 M$ at high SNR.

In summary, for the subset of parallel channels $i:b_i > e_i$, a channel inversion based power allocation based on the 
effective channel $\dfrac{b_i^2 - e_i^2}{\omega_{\nu+i}}$ will be performed, whereas for the set of parallel channels 
$j=k-r+1,\ldots,k$, a channel inversion type power allocation based on effective channel gains $1/\omega_j$ 
will be performed. A high SNR approximation of the achievable secrecy rate for the case $\t{rank}\left(\b H_e\right) < m_a$ is 
given by
\begin{equation}
R_s^{\t{high}} \approx r\,\log M + \displaystyle\sum_{i:b_i>e_i}
\left[\mathcal{I}\left(\dfrac{b_i ^2}{\omega_{\nu+i}}\, p_{\nu+i}^{\t{high}}\right) - \mathcal{I}\left(\dfrac{e_i ^2}{\omega_{\nu+i}}\, p_{\nu+i}^{\t{high}}\right)\right] .
\end{equation}
%%%%%%%%%%%%%%%%%%%%%%%%%%%%%%%%%%%%%%%%%%%%%%%%%%%%%%%%%%

%%%%%%%%%%%%%%%%%%%%%%%%%%%%%%%%%%%%%%%%%%%%%%%%%%%%%%%%%%
\section{Numerical Results}
\label{sec:numerical_result}
\subsection{Transmitter with Accurate Eavesdropper CSI}
Without any loss of generality, we assume equal noise power level at receivers of both Bob  and Eve. 
We also assume that Alice has full CSI of both Bob and Eve. Our numerical results average over 
$500$ channel realization, where each entry of both Bob's and Eve's channel matrices is i.i.d.
complex random Gaussian variable
with zero mean and unit variance.

\begin{figure}
\centering
	\includegraphics[scale=0.9]{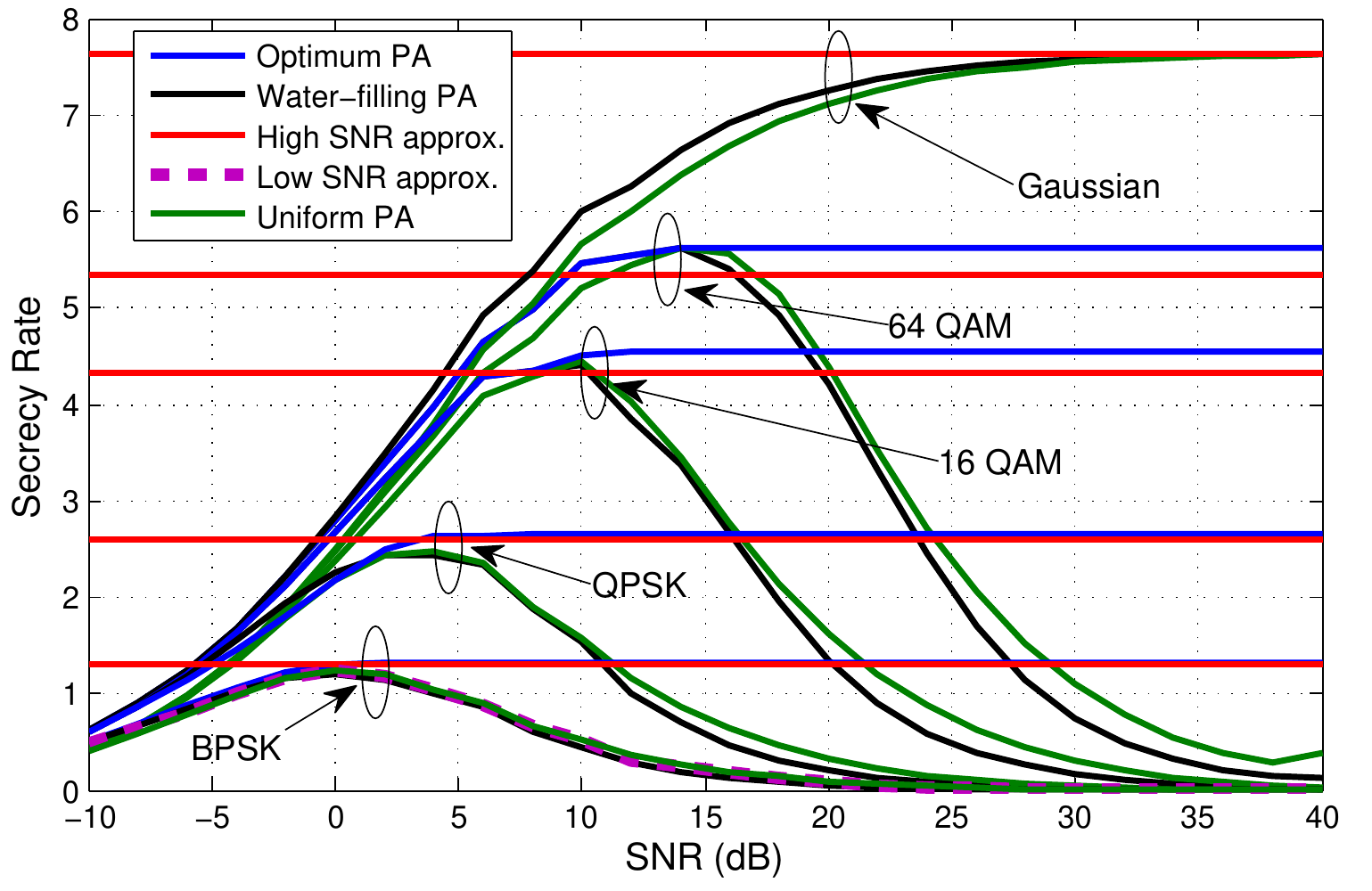}
	\caption{$5\times 5\times 5$ MIMOME system}
	\label{fig:555_mimome}
\end{figure}

In Fig. \ref{fig:555_mimome}, we present numerical test results for a $5\times 5 \times 5$ MIMOME system. 
In this test, we ensure that each realization of $\b H_e$ is non-singular, i.e., $\t{rank}(\b H_e) = 5 = m_a$. 
Since, $\t{rank}(\b H_e) = m_a$, there is no parallel channel only directed at Bob. In addition to result
for the proposed power allocation (PA) algorithm, we also present results obtained using the water-filling PA of Section 
\ref{sec:gaussian} and results from equal power over all channels (uniform PA) \cite{khisti_MIMOME}. We 
also present high SNR approximation results for every tested
constellation as well as the low SNR approximation result for BPSK. For 
other constellations (QPSK, 16-QAM and 64-QAM), low SNR approximation gives the same
result as the water-filling PA.

As seen in Fig. \ref{fig:555_mimome}, at high SNR, power allocations according to water-filling and uniform strategies would drop 
the secrecy rate to almost zero. This result is intuitive. For finite-alphabet, the achievable mutual information at high power approaches 
the saturation value of $\log_2 M$. For Gaussian input, however, the mutual information or the capacity increases monotonically with increasing power.
Since both water-filling and uniform strategies assume a Gaussian input distribution,
at high SNR both schemes would use more power to transmit signals. 
This strategy drops the secrecy rate asymptotically to zero as the difference in mutual information
between Alice-to-Bob and Alice-to-Eve narrows with at very high SNR. 
Fig. \ref{fig:555_mimome} also indicates that the high SNR approximation analysis
closely matches the secrecy rate at high SNR regime. Similarly, the secrecy rate at low SNR is also 
closely approximated by the low-SNR approximation analytical result.

\begin{figure}
\centering
	\includegraphics[scale=0.9]{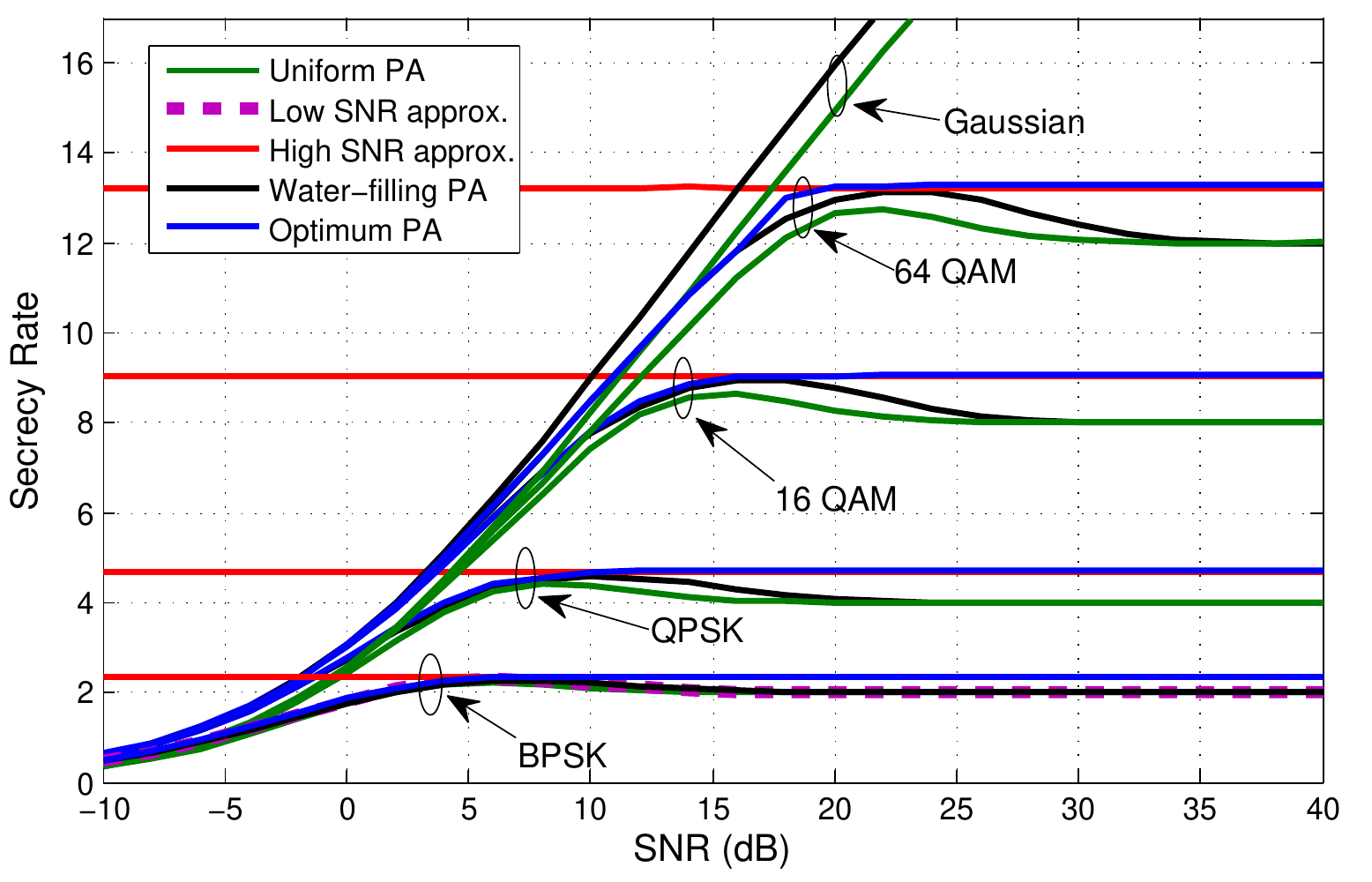}
	\caption{$5\times 5\times 3$ MIMOME system}
	\label{fig:553_mimome}
\end{figure}

Fig. \ref{fig:553_mimome} presents test results for a $5\times 5 \times 3$ MIMOME system. Because in this case $\t{rank}(\b H_e) < m_a$,
there are parallel channel components only directed towards Bob. Therefore, results obtained in this case are different from
Fig. \ref{fig:555_mimome}. Specifically, we observe that, even though
secrecy rates using water-filling and uniform PA schemes at high SNR
drops from the maximum attainable value, they do not approaches zero
as in Fig. \ref{fig:555_mimome}. The reason is due to the existence of channels
directed only at Bob. For the parallel channels with components along both at Bob's and Eve's direction, the difference in mutual information
between Alice-to-Bob and Alice-to-Eve  narrows at high SNR. However, for each of the parallel channels with components only in Bob's direction, 
the secrecy rate would approach $\log_2 M$. Therefore, at high SNR the total secrecy rate approaches $r\, log_2 M$.
Moreover, our proposed PA algorithm can achieve some additional non-zero secrecy rate from the channels directed at both Bob and Eve, in addition to 
those parallel channels only in Bob's direction.

Both Fig. \ref{fig:555_mimome} and \ref{fig:553_mimome} indicate that when using Gaussian
assumption for finite-alphabet inputs,
there is a \emph{threhold SNR} above which the achievable secrecy rate starting to decrease.
The value of this \emph{threshold SNR} is higher for the higher order constellation. Therefore, by 
adaptively switching to the next higher order modulation format once we reach the threshold point for a 
particular constellation, we will be able to use
Gaussian water-filling PA algorithm even for finite-alphabet inputs.
We also notice similar \emph{threshold SNR} for uniform PA. In addition, uniform PA also
achieves secrecy rate close to the water-filling PA. Therefore, a very simple but near optimal 
strategy would be to use uniform power allocation and
start switching to the next higher order modulation once above each
\emph{threshold SNR}.

\subsection{Transmitter with Partial Eavesdropper CSI}
Thus far, our analysis assumes that Alice possesses full channel information 
of Eve. In practice, however, Eve's CSI or even the presence of a passive Eve is difficult to determine. 
Therefore, in this section, we will numerically evaluate the scenario when 
Alice only has access to partial (statistical) information regarding Eve's channel state. In particular, let Eve's channel consists of
\[
\b H_e = \hat{\b H}_e + \b E_e .
\]
Here, $\hat{\b H}_e$ is Eve's mean CSI known to Alice whereas $\b E_e$ is the CSI uncertainty which 
is modeled as zero mean white Gaussian noise with variance
$\sigma_e^2$, i.e., $\b E_e\sim\mathcal{CN}\left(\b 0,\sigma_e^2 \b I\right)$.

In this case, if Alice performs power allocation based on the known observation $\hat{\b H}_e$ by disregarding 
the uncertainty, the power allocation may not be optimal. In fact, Alice may 
even lose secrecy. In the following proposition, 
we present the achievable ergodic secrecy rate for a given power allocation $\b p$.

\begin{proposition}
\label{prop:partial_eve}
For a given power allocation $\b p$ based on the known observation $\hat{\b H}_e$ without considering the uncertainty, the achievable ergodic secrecy rate $R_{\t{sec}} ( \b p)$ can be given as follows
\begin{eqnarray}
R_{\t{sec}} ( \b p) &=& \displaystyle\sum_{i:b_i > e_i} \left[ \mathcal{I} \left( \frac{b_i^2}{\omega_{\nu+i}} p_{\nu+i} \right) - \mathbb{E}\left\{\mathcal{I}\left(\dfrac{\left(e_{i}^{2}+\tilde{e}_{\nu+i}\right)^{2}}{\sigma_{\nu+i}\omega_{\nu+i}}p_{\nu+i}\right)\right\} \right] \nonumber \\
&& \qquad\qquad\qquad\qquad\qquad+
\displaystyle \sum_{j=k-r+1}^k \left[ \mathcal{I} \left( \dfrac{1}{\omega_j} p_j \right) - \mathbb{E}\left\{\mathcal{I}\left(\dfrac{\tilde{e}_{j}^{2}}{\sigma_{j}\omega_{j}}p_{j}\right)\right\} \right]
\label{eq:unknown_eve}
\end{eqnarray}
Here, $\tilde{e}_{i} \sim \mathcal{CN} \left(0, \sigma_e^2\,\omega_i \right)$, $i=1,\ldots,k$ are i.i.d. random. Also, $\sigma_i^2 = 1 + \sigma_e^2\, \displaystyle \left( \sum_{\ell = 1}^k p_\ell - p_i \right)$.
\end{proposition}

\begin{proof}
See Appendix \ref{app:partial_eve}.
\end{proof}

From Eq. \eqref{eq:unknown_eve} we notice that, data symbols $s_{k-r+1},\ldots,s_k$ - previously only observed 
by Bob, are also seen by Eve now. In addition, for the symbols $s_{\nu+i}, \ldots, s_{\nu + s}$ - observed 
by both Bob and Eve, an additional uncertainty term $\tilde{e}_{\nu+i}$ is added to the mutual information of Eve.

\begin{figure}
\centering
	\includegraphics[scale=0.9]{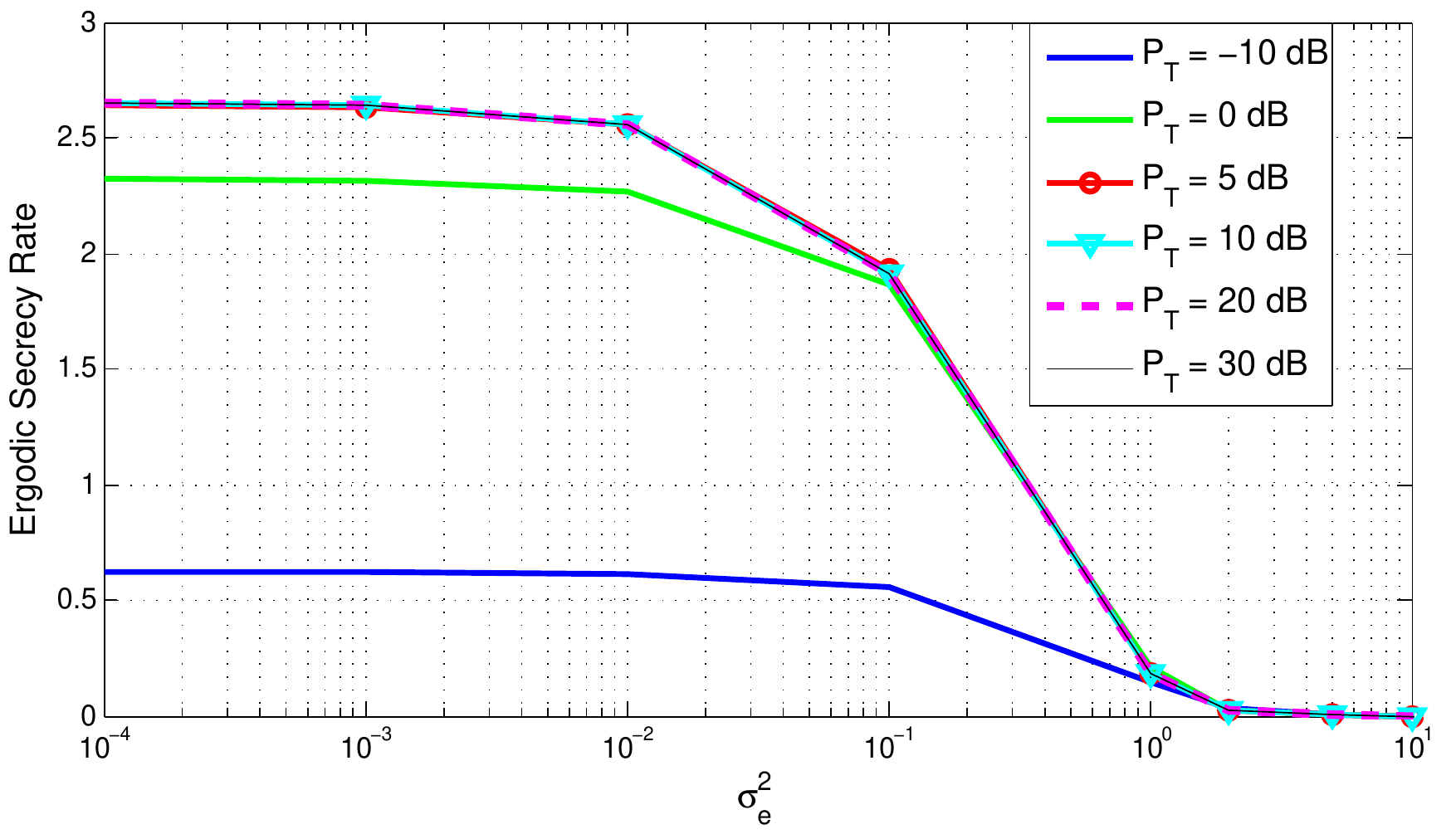}
	\caption{Secrecy Rate of a $5\times 5\times 5$ MIMOME system with partial Eve's CSI}
	\label{fig:555_mimome_avgCSI}
\end{figure}

\begin{figure}
\centering
	\includegraphics[scale=0.9]{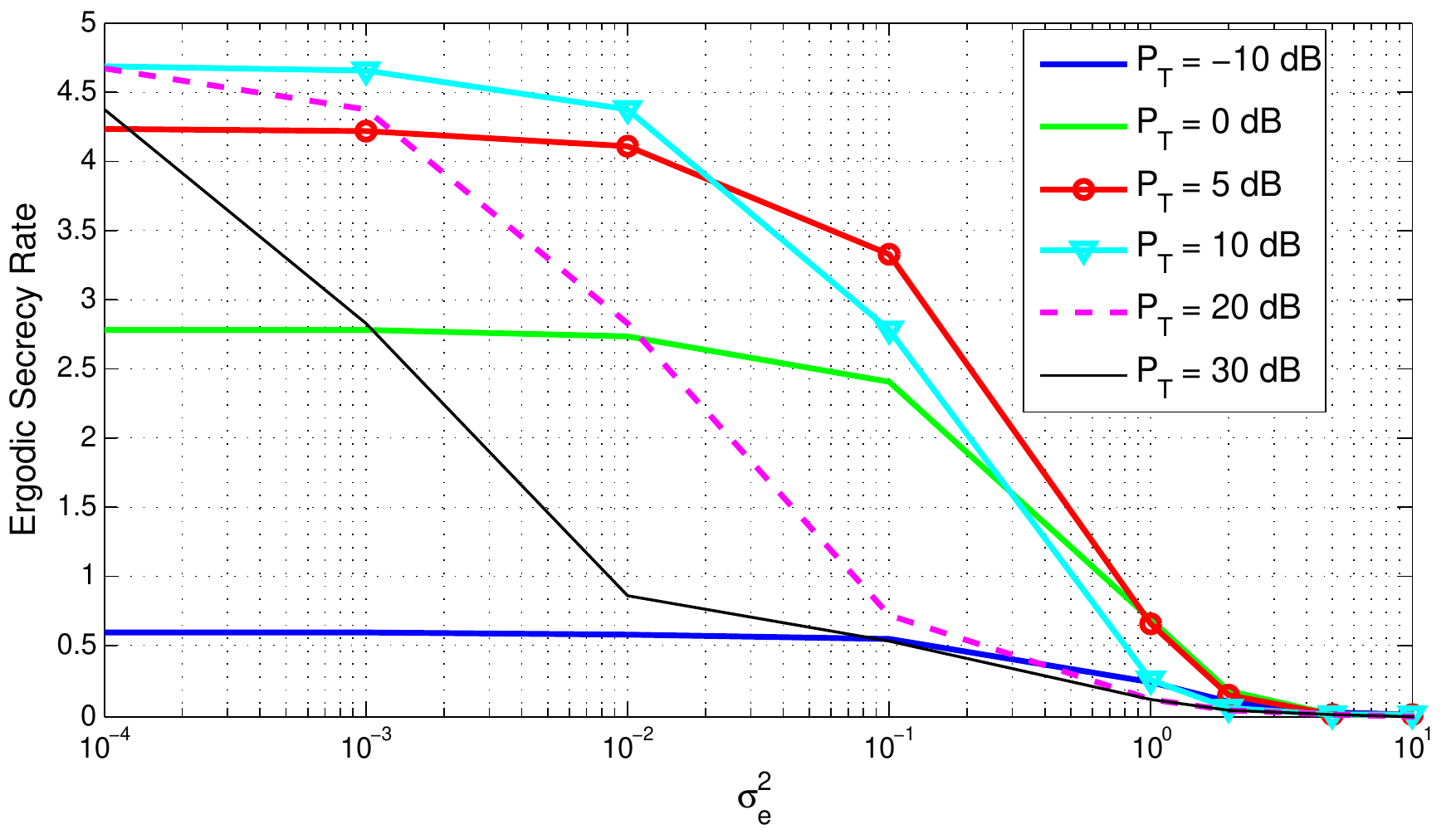}
	\caption{Secrecy Rate of a $5\times 5\times 3$ MIMOME system with partial Eve's CSI}
	\label{fig:553_mimome_avgCSI}
\end{figure}

In Fig. \ref{fig:555_mimome_avgCSI} and \ref{fig:553_mimome_avgCSI}, we present test results of 
achievable ergodic secrecy rate for different values of variance of channel uncertainty, for 
$5\times 5\times 5$ and $5\times 5\times 3$ MISOME systems, 
respectively. Here, Alice is only aware of the mean CSI $\hat{\b H}_e$ and use this CSI for power control without considering the 
uncertainty component $\b E_e$. We present test results for different values of available transmission power. 
Both Fig. \ref{fig:555_mimome_avgCSI} and \ref{fig:553_mimome_avgCSI} indicate that the achievable ergodic secrecy rate
decreases with larger channel uncertainty. In addition, we observe from Fig. \ref{fig:555_mimome_avgCSI} that the achievable ergodic secrecy rate 
plots for  transmission power $P_T\geq 5$ stays the same.
This is because Alice does not require all available power to transmit at higher SNR.
Hence, the power allocation stays the same at higher SNR. 
However, in Fig.~\ref{fig:553_mimome_avgCSI}, we observe that the achievable ergodic secrecy rate in fact decreases with larger transmission power
for $P_T > 10$ dB. In this case, $\t{rank}\left(\hat{\b H}_e\right) < m_a$. When Alice only uses this information for power control without 
considering the uncertainty, Alice will likely allocate more power to the bank of parallel channels which only has components toward Bob. 
At high SNR, Alice will allocate more powers to these channels. However, as shown in Eq. \eqref{eq:unknown_eve}, because of channel uncertainty 
$\b E_e$, Eve now also possesses components along these channels. In other words, Eve can also receive signals from these
channels. As SNR grows large, the mutual information for finite alphabet saturates to $\log_2 M$. 
As a result, Eve can receive nearly full data information in these channel and consequently, the achievable ergodic secrecy rate will decrease at high SNR.

We note that, in order to maximize the achievable ergodic secrecy rate under channel uncertainty, one needs to solve the following optimization problem
\begin{equation}
\begin{split}
&\underset{\b p}{\t{maximize}} ~ \displaystyle\sum_{i = 1}^s \left[ \mathcal{I} \left( \frac{b_i^2}{\omega_{\nu+i}} p_{\nu+i} \right) - \mathbb{E}\left\{\mathcal{I}\left(\dfrac{\left(e_{i}^{2}+\tilde{e}_{\nu+i}\right)^{2}}{\sigma_{\nu+i}\omega_{\nu+i}}p_{\nu+i}\right)\right\} \right] \nonumber \\
& \qquad\qquad\qquad\qquad\qquad\qquad\qquad\qquad+
\displaystyle \sum_{j=k-r+1}^k \left[ \mathcal{I} \left( \dfrac{1}{\omega_j} p_j \right) - \mathbb{E}\left\{\mathcal{I}\left(\dfrac{\tilde{e}_{j}^{2}}{\sigma_{j}\omega_{j}}p_{j}\right)\right\} \right]\\
&\t{subject\, to\,:}~~~~
\displaystyle\sum_{i=1}^s p_{\nu+i} + \displaystyle\sum_{j=k-r+1}^{k} p_j \leq P_T.
\end{split}
\end{equation}
Investigation of this optimization problem is beyond the scope of this work. However, in \cite{basharsecrecy}, 
a power allocation algorithm to maximize the achievable ergodic secrecy rate 
under arbitrary input distribution is provided for a multiple-input single-output and single-eavesdropper scenario. 
An extension of the power allocation algorithm provided in \cite{basharsecrecy} can also be used to solve the above optimization problem.

%%%%%%%%%%%%%%%%%%%%%%%%%%%%%%%%%%%%%%%%%%%%%%%%%%%%%%%%%%

%%%%%%%%%%%%%%%%%%%%%%%%%%%%%%%%%%%%%%%%%%%%%%%%%%%%%%%%%%
\section{Conclusion}
\label{sec:conclusion}
This work considers the effect of practical finite-alphabet inputs on the secrecy performance of an MIMOME system. 
Our investigation led to the application of a precoding matrix to convert the MIMOME system into a bank of parallel channels
so as to reformulate the achievable secrecy rate problem. We proposed a decentralized dual decomposition and a
corresponding power allocation  algorithm to maximize the achievable secrecy rate based on the proposed precoding for
channel transformation. 
We analyzed the Gaussian input as a special case and provided a water-filling inspired power allocation strategy. 
Furthermore, we derived analytical results of achievable secrecy rate based on approximations
at low and high SNR scenarios. Our results show that
power allocation strategy based on Gaussian input is far from 
optimal when applied blindly in finite-alphabet input situation and may in fact be
very risky by driving the secrecy rate to zero at higher SNR.
%%%%%%%%%%%%%%%%%%%%%%%%%%%%%%%%%%%%%%%%%%%%%%%%%%%%%%%%%%

%%%%%%%%%%%%%%%%%%%%%%%%%%%%%%%%%%%%%%%%%%%%%%%%%%%%%%%%%%
\appendices
\section{Proof of Proposition \ref{thm:secrecy}}
\label{app:theorem}
An achievable secrecy rate for an arbitrary distribution of input $\b x$ can be obtained by solving the following optimization problem

\begin{equation}
\begin{split}
&\underset{\b K_{\b x}}{\t{maximize}}~I\left(\b x;\b y_{b}\right)-I\left(\b x;\b y_{e}\right)\\
&\t{subject\, to\,:}~~~~\b K_{\b x}\succeq0,\,\,{\b{K_{\b x}=\b K_{\b x}^{H}}}\\
&\phantom{\t{subject\, to\,:}~~~~}\t{tr}\left(\b K_{\b x}\right)\leq P_{T}
\end{split}
\label{eq:main_problem_modified}
\end{equation}
When we apply the precoding matrix $\b W$ from Eq. \eqref{eq:precoding_matrix},
the input of the system becomes $\b x = \b W \b s$.
The objective function of the above optimization problem can be written as follows

\begin{align}
& I\left(\b x;\b y_{b}\right)-I\left(\b x;\b y_{e}\right)\\
=& I\left(\b s;\b H_{b}\b W \b s+\b n_{b}\right)-I\left(\b s;\b H_{e}\b W \b s+\b n_{e}\right)
\label{eq:line2}\\
=& I\left(\b s;\tilde{\b H}_{b} \b s+\tilde{\b n}_{b}\right)-I\left(\b s;\tilde{\b H}_{e} \b s+\tilde{\b n}_{e}\right)
\label{eq:line3}\\
=& \left[\displaystyle\sum_{i=1}^s \mathcal{I}\left(b_i^2\, p_{k-r-s+i} \right) + \displaystyle\sum_{j=k-r+1}^k \mathcal{I} \left(p_j\right)\right] - \left[\displaystyle\sum_{\ell=1}^{k-r-s} \mathcal{I}\left(p_{\ell}\right) + \displaystyle\sum_{i=1}^s \mathcal{I}\left(e_i^2\, p_{k-r-s+i}\right)	\right]
 \label{eq:line4}\\
=& \displaystyle\sum_{j=k-r+1}^k \mathcal{I} \left(p_j\right) + \displaystyle\sum_{i=1}^s \left[\mathcal{I}\left(b_i^2\, p_{k-r-s+i} \right) -\mathcal{I}\left(e_i^2\, p_{k-r-s+i}\right)\right] - \displaystyle\sum_{\ell=1}^{k-r-s} \mathcal{I}\left(p_{\ell}\right).
\end{align}
Note that \eqref{eq:line3} follows from \eqref{eq:line2} since linear unitary transformation of
channel outputs preserves mutual information.
Because \eqref{eq:line3} represents the difference in mutual information between a subset of parallel channels,
we can rewrite \eqref{eq:line3} into the summation form of \eqref{eq:line4}.

Since $\b s$ is a random vector with identity correlation matrix, $\b K_{\b x}=\mathbb{E}\left[\b x\b x^{H}\right]=\b W\b W^{H}$, we have 
$\t{tr}\left(\b K_{\b x}\right)=\sum_{i=1}^{k}\omega_{i}p_{i}$. Therefore, we can reformulate the 
optimization problem in \eqref{eq:main_problem_modified} as follows

\begin{equation}
\begin{split}
&\underset{\{p_i\}}{\t{maximize}}
  \sum_{i=1}^s \left[\mathcal{I}\left(b_i^2\, p_{k-r-s+i} \right) -\mathcal{I}\left(e_i^2\, p_{k-r-s+i}\right)\right] - \sum_{i=1}^{k-r-s} \mathcal{I}\left(p_{i}\right)
+\sum_{i=k-r+1}^k \mathcal{I} \left(p_i\right) \\
&\t{subject\, to\,:}~~~~~~~~~~~\sum_{i=1}^{k}\omega_{i}p_{i}\leq P_T
\end{split}
\label{eq:main_problem_parallel}
\end{equation}

\begin{figure}
\centering
	\includegraphics[scale=0.85]{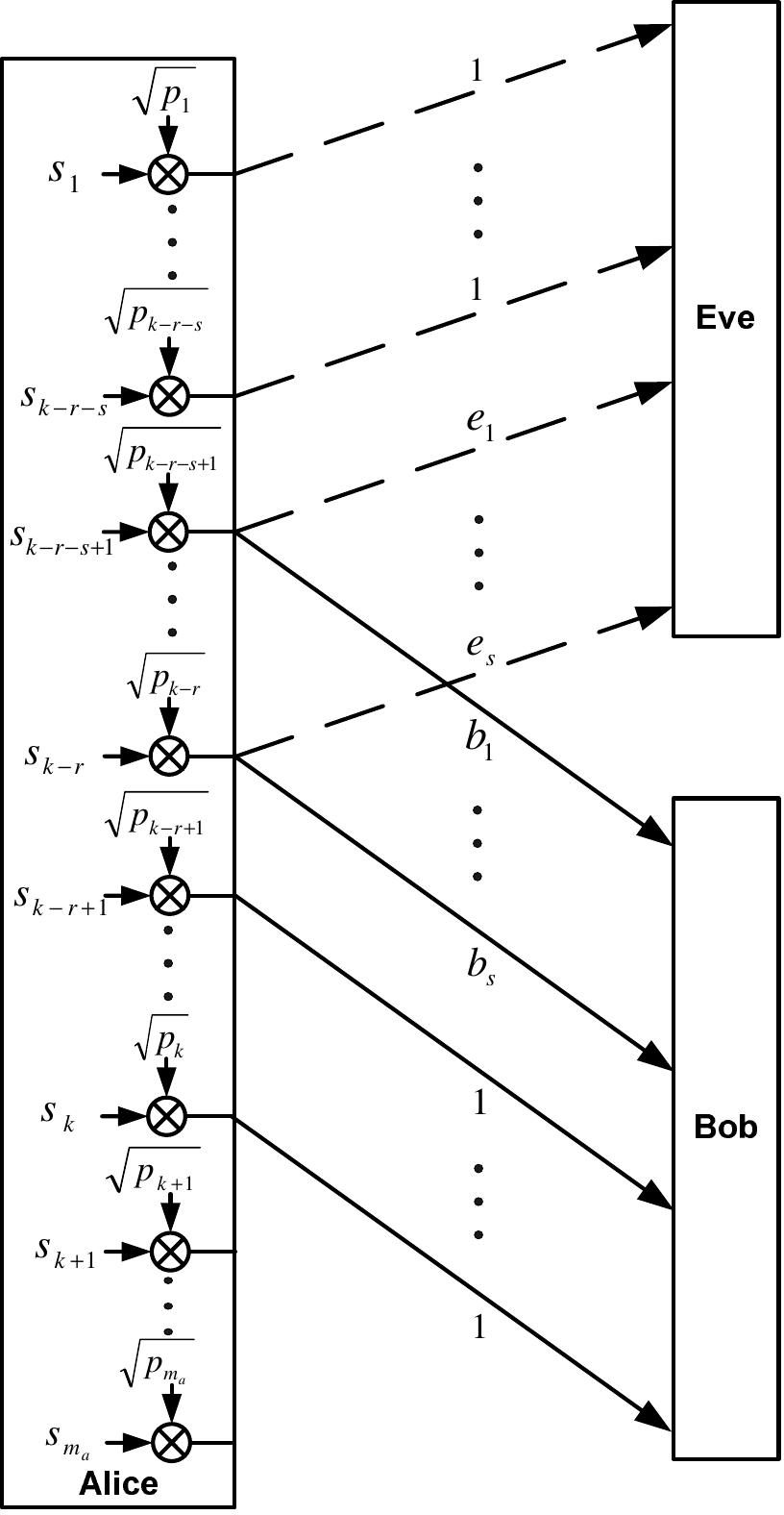}
	\caption{Precoding matrix $\b W = \boldsymbol{\Psi}_{a}\b B\b P^{\nicefrac{1}{2}}$ converts
	MIMOME channel to a bank of parallel channels}
	\label{fig:mimome_to_parallel}
\end{figure}

Note that the above optimization problem is a distributed power allocation problem for the bank
of parallel channels shown in Fig. \ref{fig:mimome_to_parallel},  where the variables $\{ p_i \}$'s are coupled through the total power constraint.
Because $\mathcal{I}(.)\geq 0$, we have
optimum power allocation $p_i^* = 0$ for $i=1,\ldots, k-r-s $.  In addition,
since $\mathcal{I}(\gamma)$ are monotonically increasing in
$\gamma$, $\mathcal{I}\left(b_i^2\, p_{k-r-s+i} \right)-\mathcal{I}\left(e_i^2\, p_{k-r-s+i}\right)\leq 0$
whenever $b_i\leq e_i$, hence $p_i^* = 0$, for all $i$ with $b_i\leq e_i$.
 Furthermore, from Fig. \ref{fig:mimome_to_parallel}, we observe that symbols $s_{k+1},\ldots,s_{m_a}$ are
transmitted towards the direction of $\mathcal{S}_n$ and is not part of the optimization problem \eqref{eq:main_problem_parallel}.
Hence, it will be wasteful to expend any transmission power to transmit
these symbols. Therefore, $p_i^*=0$ for $i=k+1,\ldots,m_a$. By change of variables, $p'=\omega_i\,p_i$, and then replacing $p'$ with $p$, we obtain the optimization
problem in Eq. \eqref{eq:optimization_parallel}.

\section{Proof of Proposition \ref{prop:sufficient_condition}}
\label{app:sufficient_condition}
If $b_i^2 \leq e_i^2$, the difference in mutual information between Alice-to-Bob and Alice-to-Eve 
channel will be negative. Hence, no secrecy is possible. As a result, the optimal power allocation should be zero.

When $b_i^2 > e_i^2$, it is possible to achieve a positive secrecy rate. If, in the solution of the original 
problem \eqref{eq:optimization_parallel}, the sum power constraint inequality becomes a strict inequality, 
then $\mu = 0$ according to the KKT condition \cite{boyd2004convex}. Such condition can occur at high SNR regime. 
Therefore, the solution of problem \eqref{eq:subproblem1_sol} will be achieved by the power allocation $p'$ that renders the 
MMSE difference function in Eq. \eqref{eq:mmseD} zero. If the
MMSE difference function admits a unique zero solution, then the solution of 
Eq. \eqref{eq:subproblem1_sol} will be optimal, i.e., $p_{\nu +i}^* = p'$.

If on the other hand, the sum power constraint in the original problem admits an equality, then $\mu > 0$. If the 
MMSE difference function is monotonically decreasing for $0\leq p\leq p'$, then the function $f_{\t{mmseD}} (p,b_i,e_i,\omega_{\nu + i}) - \mu$ will be
zero for a power allocation value $p^*$ that is less than $p'$. Now, if MMSE 
difference function is strictly monotonically decreasing, then the power allocation solution will be unique. 
Therefore, the solution of Eq. \eqref{eq:subproblem1_sol} will be optimal.

\section{Proof of Proposition \ref{prop:unique_zero}}
\label{app:unique_zero}
Define, $\rho_1 = b_i\,p$ and $\rho_2 = e_i\,p$. First, we will prove that the difference function, $g_{\t{D}}(\rho_1,\rho_2) = g(\rho_1) - g(\rho_2)$ admits a unique zero for $p>0$. 

Assume that the strictly unimodal function $g(\rho)$ is strictly monotonically increasing for $\rho \leq m$ and strictly monotonically decreasing for $\rho > m$. 
When, $\rho_2<\rho_1\le m$, the difference function $g_{\t{D}}(\rho_1,\rho_2)$ cannot be zero 
due to the strictly monotonically increasing 
nature of the function $g(\rho)$. 
Similarly, when $\rho_1 > \rho_2 > m$, the difference function $g_{\t{D}}(\rho_1,\rho_2)$ cannot be zero 
due to the strictly monotonically decreasing nature of the function $g(\rho)$. Therefore,  
the difference function $g_{\t{D}}(\rho_1,\rho_2)$ can be zero {\em only } when $\rho_2 < m < \rho_1$. 

Suppose, for $p = p'$, the difference function $g_{\t{D}}(b_i\,p,e_i\,p')$ equals zero. For $p>p'$, $g(b_i\, p) > g(b_i\, p')$ and $g(e_i\, p) < g(e_i\, p')$ 
due to the strictly unimodal properties of $g(\rho)$. Hence, when $\rho_2 < m < \rho_1$,
the difference function $g_{\t{D}}(\rho_1,\rho_2) = g(\rho_1) - g(\rho_2)$ cannot be zero for $p>p'$. 
The same argument holds for any point $p<p'$. Therefore, when $b_i > e_i$, the difference function  $g_{\t{D}}(\rho_1,\rho_2)$ admits a unique 
zero for $p>0$. Hence the MMSE difference function $f_{\t{mmseD}} (p,b_i,e_i,\omega_{\nu + i})$ has a unique zero solution for $p >0$

\section{Proof of Proposition \ref{prop:partial_eve}}
\label{app:partial_eve}
With $\b H_e = \hat{\b H}_e + \b E_e$, Eq. \eqref{eq:eve_parallel} can be written as
\begin{equation}
\tilde{\b y}_{e}=\tilde{\b H}_{e}\,\b s + \boldsymbol{\Psi}_{e}^{H}\b E_{e}\b W\b s +\tilde{\b n}_{e}
\end{equation}
Here, $\tilde{\b H}_{e} = \boldsymbol{\Psi}_{e}^{H}\hat{\b H}_{e}\b W = \boldsymbol{\Sigma}_{e}\left[\begin{array}{cc}
\b I & \b 0\end{array}\right]\b P^{\nicefrac{1}{2}}$. Now, the second term in the above equation can be expressed as $\boldsymbol{\Psi}_{e}^{H}\b E_{e}\b W\b s = \tilde{\b E}_{e}\b P^{\nicefrac{1}{2}}\b s$, where, $\tilde{\b E}_{e} = \hat{\b E}_{e}\b B$ and $\hat{\b E}_{e} =	\boldsymbol{\Psi}_{e}^{H}\b E_{e}\boldsymbol{\Psi}_{a}$. We can decompose $\tilde{\b E}_{e}$ as a sum of two matrices, where the first matrix is a diagonal matrix containing the diagonal elements of $\tilde{\b E}_{e}$. The second matrix contains the non-diagonal entries of $\tilde{\b E}_{e}$ and contains all zero elements in its diagonal. The decomposition is given below

\begin{equation}
\tilde{\b E}_{e} =
\left[\begin{array}{ccccccc}
\tilde{e}_{1} & 0 & \ldots & 0 & 0 & \ldots & 0\\
0 & \tilde{e}_{2} & \ddots & 0 & 0 & \ddots & 0\\
\vdots & \vdots & \ddots & \vdots & \vdots & \ddots & \vdots\\
0 & 0 & \ldots & \tilde{e}_{k} & 0 & \ldots & 0\\
0 & 0 & \ldots & 0 & 0 & \ldots & 0\\
\vdots & \vdots & \ddots & \vdots & \vdots & \ddots & \vdots\\
0 & 0 & \ldots & 0 & 0 & \ldots & 0\end{array}\right] +\left[\begin{array}{ccccccc}
0 & \tilde{e}_{12} & \ldots & \tilde{e}_{1k} & 0 & \ldots & 0\\
\tilde{e}_{21} & 0 & \ddots & \tilde{e}_{2k} & 0 & \ddots & 0\\
\vdots & \vdots &  & \vdots & \ddots & \vdots\\
\tilde{e}_{k1} & \tilde{e}_{k2} & \ldots & 0 & 0 & \ldots & 0\\
\tilde{e}_{(k+1)1} & \tilde{e}_{(k+1)2} & \ldots & \tilde{e}_{\left(k+1\right)k} & 0 & \ldots & 0\\
\vdots & \vdots & \ddots & \vdots & \vdots & \ddots & \vdots\\
\tilde{e}_{m_{e}1} & \tilde{e}_{m_{e}2} & \ldots & \tilde{e}_{m_{e}k} & 0 & \ldots & 0\end{array}\right]
\end{equation}

Here, $\tilde{e}_{ij} \sim \mathcal{CN}\left( 0,\sigma_e^2 \omega_j \right)$. Based on the above decomposition and replacing $p_i$ with $\dfrac{p_i}{\omega_i}$, the $i$-th entry of the vector $\tilde{\b y}_{e}$ can be expressed as follows

\begin{equation}
\tilde{y}_{e_i} = \left( e_i' + \tilde{e}_i \right) \dfrac{1}{\sqrt{\omega_i}} \sqrt{p_i} s_i + g_i + n_{e_i}
\end{equation}

Here, $g_i = \displaystyle\sum_{\ell=1,\ell\neq i}^k \tilde{e}_{i\ell} \dfrac{1}{\sqrt{\omega_\ell}} \sqrt{p_\ell} s_\ell$ and $e_i'$ can be expressed as

\begin{equation}
e_{i}'=\begin{cases}
1, & i=1,\ldots,k-r-s\\
e_{i-\nu}, & i=\nu+1,\ldots,\nu+s\,\t{and\,}\nu=k-r-s\\
0, & i=k-r+1,\ldots,m_{e}\end{cases}
\end{equation}

Assume that the eavesdropper is performing conventional decoding on each of these parallel branch by considering the cross terms as a part of noise. Then the achievable ergodic secrecy rate for a given power allocation $\b p$ can be written as in Eq. \eqref{eq:unknown_eve}. If in case, Eve employs advanced decoding scheme, e.g. successive interference cancellation etc. then Eq. \eqref{eq:unknown_eve} will serve as a upper bound on the achievable ergodic secrecy rate.

\bibliographystyle{IEEEtran}
\bibliography{bib_security}

\newpage

\newpage

\newpage

\begin{figure}
\centering
	\includegraphics[scale=0.9]{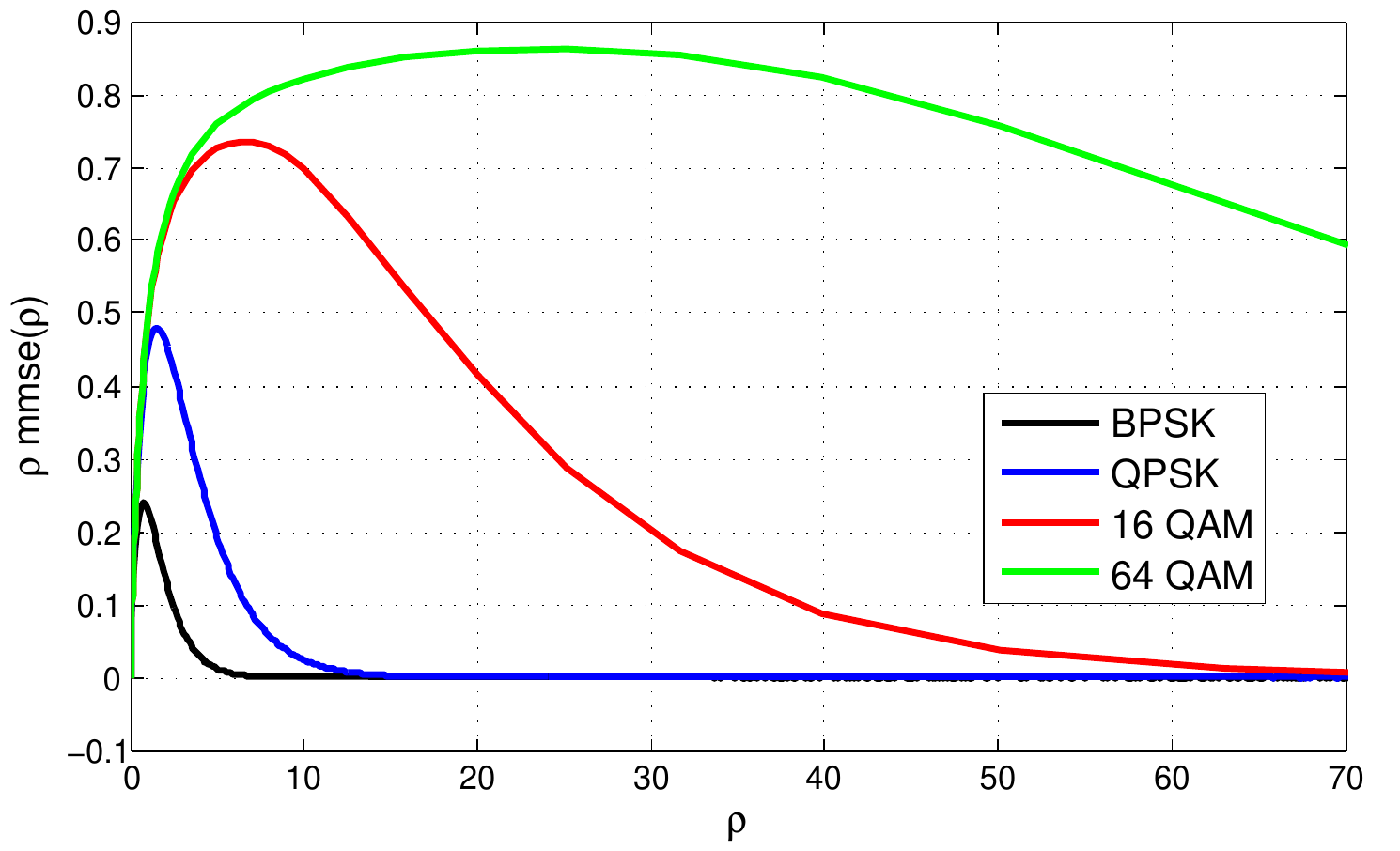}
	\caption{$\rho$ vs, $\rho\,\t{mmse}(\rho)$ plot for BPSK, QPSK, $16$-QAM and $64$-QAM constellations.}
	\label{fig:rhommse}
\end{figure}

\newpage

\newpage

\newpage

\newpage

\end{document}